\documentclass[lettersize,journal]{IEEEtran}
\usepackage{amsmath,amsfonts}
\usepackage{array}
\usepackage[caption=false,font=normalsize,labelfont=sf,textfont=sf]{subfig}
\usepackage{textcomp}
\usepackage{stfloats}
\usepackage{url}
\usepackage{verbatim}
\usepackage{graphicx}
\usepackage{pifont}
\usepackage{marvosym}
\usepackage{balance}
\usepackage[ruled,vlined]{algorithm2e}

\usepackage{makecell}
\usepackage{booktabs} 
\usepackage{multirow}
\usepackage{url}
\usepackage[backref]{hyperref}
\usepackage[table]{xcolor}
\usepackage{xcolor}
\usepackage{marvosym}
\usepackage{tcolorbox}

\newtcolorbox[auto counter, number within=section]{mybox}[2][]{colback=gray!10!white, colframe=gray!70!black, boxrule=0.8mm, width=\linewidth, arc=1mm, rightrule=0mm, toprule=0mm, bottomrule=0mm, leftrule=1mm, #1}

\def\BibTeX{{\rm B\kern-.05em{\sc i\kern-.025em b}\kern-.08em
    T\kern-.1667em\lower.7ex\hbox{E}\kern-.125emX}}

\begin{document}
\title{UDFS: Lightweight Representation-Driven Open World Robust Encrypted Traffic Classification}
\author{Youquan Xian$^*$, Xueying Zeng$^*$, Aoxiang Zhou, Jinqiao Shi, Zhiyu Hao,  Lei Cui\textsuperscript{\Letter}, Peng Liu\textsuperscript{\Letter}

\thanks{The research was supported in part by the National Natural Science Foundation of China (Nos. 62166004, U21A20474, 62262003), the Guangxi Science and Technology Major Project (No.AA22068070). $^*$Youquan Xian and Xueying Zeng have made equal contributions to this research. \textsuperscript{\Letter} Lei Cui and Peng Liu are the corresponding authors.}

\thanks{Youquan Xian and Jinqiao Shi are with the School of Cyberspace Security, Beijing University of Posts and Telecommunications, Beijing, 100876, China (e-mail: xianyq@bupt.edu.cn, shijinqiao@bupt.edu.cn).}

\thanks{Xueying Zeng is with the School of Computer Science and Engineering, Beihang University, Beijing, 100191, China (e-mail: xyz502@buaa.edu.cn).}

\thanks{Aoxiang Zhou and Peng Liu are with the School of Computer Science and Engineering, Guangxi Normal University, Guilin, 541004, China (e-mail: soar@stu.gxnu.edu.cn, liupeng@gxnu.edu.cn).}

\thanks{Zhiyu Hao is with the Zhongguancun Laboratory, Beijing, China (e-mail: haozy@zgclab.edu.cn).}

\thanks{Lei Cui is with the Zhongguancun Laboratory, Beijing, China, and also with the School of Cyberspace Science, Harbin Institute of Technology, Harbin, 150001, China (e-mail: cuilei@zgclab.edu.cn).}

}

\markboth{Journal of \LaTeX\ Class Files,~Vol.~18, No.~9, September~2020}%
{How to Use the IEEEtran \LaTeX \ Templates}

\maketitle

\begin{abstract}
In encrypted traffic analysis, sequence features such as packet length sequences have become a key basis for enabling effective classification by deep learning models. However, existing approaches to utilizing sequence features face a precision–overhead dilemma: flow-level methods lack sufficient representational capacity, making it difficult to distinguish fine-grained applications, whereas trace-level methods, although achieving higher accuracy, incur substantial computational and storage overheads, hindering their practical deployment. In addition, current open-world evaluation methods generally rely on fixed global decision strategies, overlooking inherent differences in inter-class discriminability, limiting their classification capability.
To address these challenges, this paper proposes a lightweight method for representing encrypted traffic, known as the Up-Down Flow Sequence (UDFS). It abstracts each flow within a trace as the sum of its upstream and downstream traffic, achieving an order-of-magnitude reduction in feature length while preserving key behavioral patterns, effectively balancing representational ability and computational cost. Building on this representation, we further develop a prototype-based adaptive open-world traffic recognition mechanism. This mechanism learns personalized decision boundaries for each category, thereby enhancing classification performance. Extensive experiments across multiple challenging scenarios, including coarse- and fine-grained classification, concept drift, and open-world conditions, demonstrate that our proposed method significantly outperforms existing approaches. The code and datasets are available at \url{https://github.com/kid1999/UDFS}.
\end{abstract}


\section{Introduction}
\begin{figure}[t]
\centering
\subfloat[\small High redundancy in flow sequences]{\label{fig:flow_repetition}\includegraphics[width=0.45\linewidth]{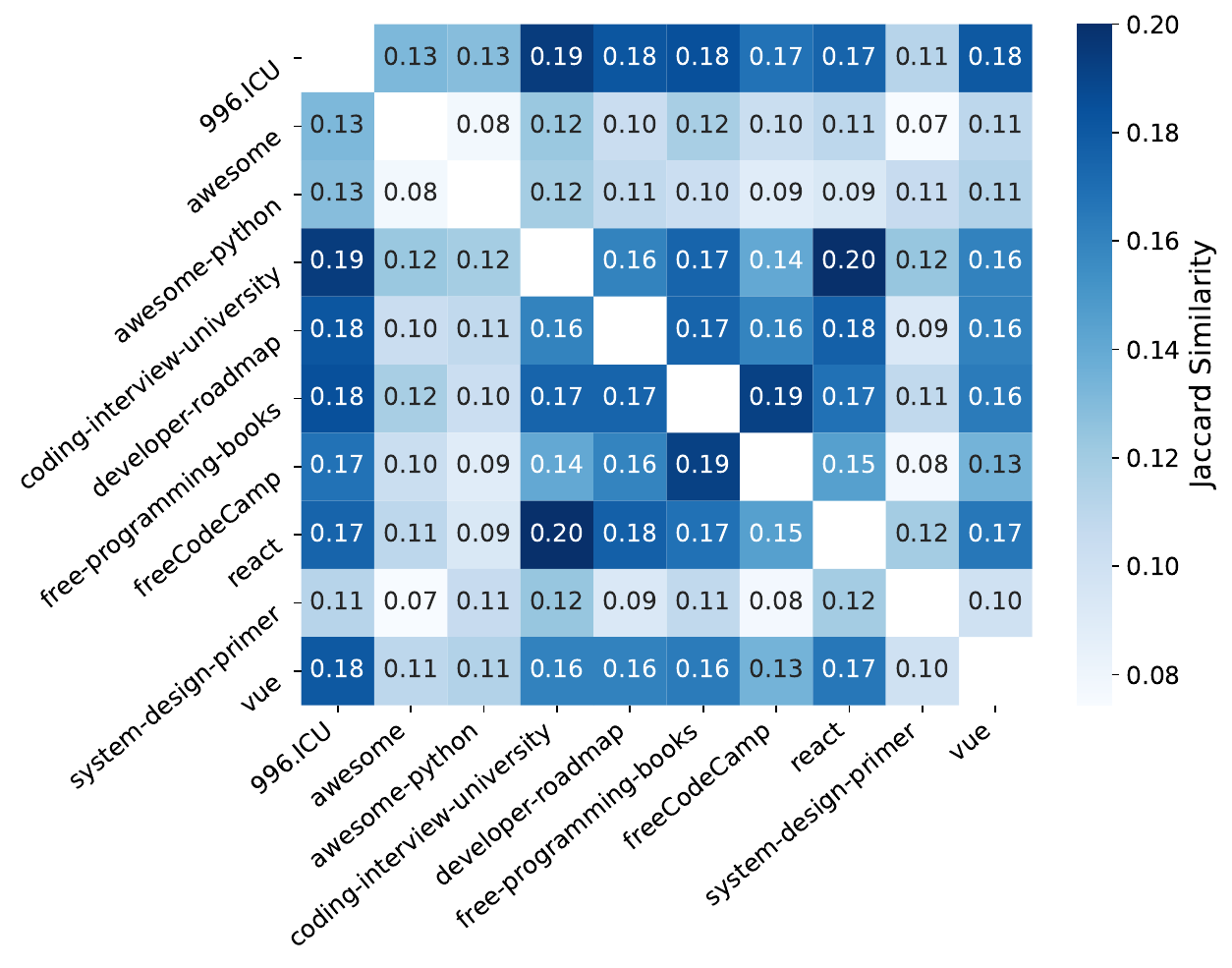}}\quad
\subfloat[\small Long trace sequences]{\label{fig:packet_count}\includegraphics[width=0.45\linewidth]{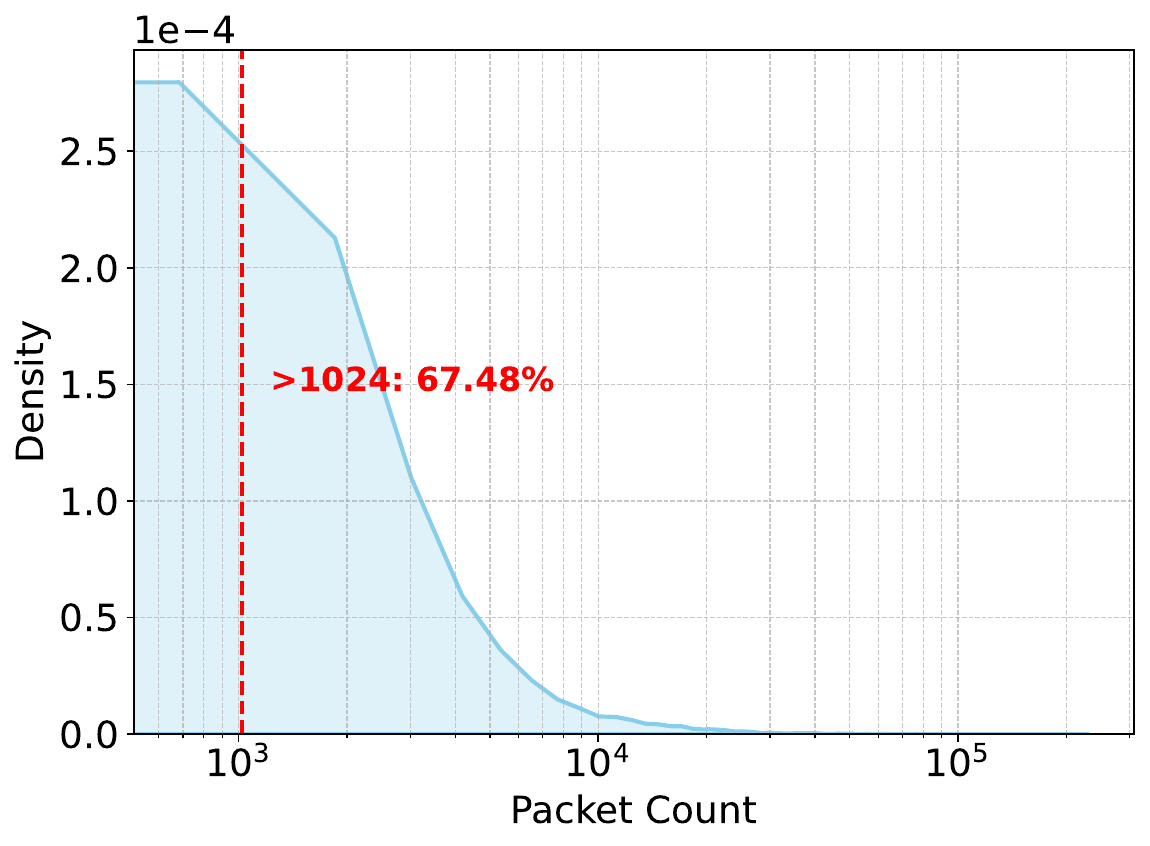}}\\
\caption{Flow sequence Jaccard similarity (10 GitHub repositories) and trace sequence length distribution (10,000 websites).}
\label{fig:start}
\vspace{-5mm}
\end{figure}

\IEEEPARstart{W}{ith} 
the widespread adoption of network encryption technologies, traditional traffic identification methods that rely on plaintext content have gradually become ineffective. Consequently, Encrypted Traffic Analysis (ETA) has emerged as a key technology for network security and service optimization \cite{sharma2025survey,papadogiannaki2021survey}. In deep learning–driven ETA approaches, the selection of input features plays a decisive role in determining both the discriminative power and generalization ability of the model \cite{shen2022machine}. Among various feature types—including statistical features, raw byte sequences, and packet sequences—the packet sequence has demonstrated distinct advantages in characterizing traffic behavior \cite{chen2022length,liu2019fs,taylor2017robust}. Compared to statistical features, packet sequences avoid the need for manual feature extraction based on expert knowledge and prevent the information loss associated with such processes; compared to raw byte sequences, they significantly reduce data dimensionality while mitigating the risk of the model learning spurious correlations \cite{wickramasinghe2025sok,zhao2025sweet}. These advantages make packet sequences an essential feature in contemporary deep learning–based encrypted traffic analysis.

Despite the clear advantages of packet sequences, effectively modeling them remains a core challenge. Existing strategies have mainly evolved along two directions: flow-level and trace-level. Flow-level methods, e.g., FS-Net \cite{liu2019fs}, GGFAST \cite{piet2023ggfast}, FlowPic \cite{shapira2021flowpic}, treat individual flows as the basic unit of analysis. These methods are lightweight and easy to deploy. However, flow-level features may exhibit similar traffic patterns across different applications, as shown in Fig. \ref{fig:flow_repetition}, making it difficult for models to capture critical differences. This limitation reduces their effectiveness in distinguishing between behaviorally similar applications.
Trace-level methods, e.g., Deep Fingerprinting \cite{sirinam2018deep}, RLTree \cite{li2021rltree}, TSHC-SW \cite{chen2023classify}, Qu et al. \cite{qu2023input}, attempt to model multiple flows or entire traces as a whole to obtain richer contextual and behavioral information. Although this approach offers potential advantages in accuracy, its input sequences are often extremely long, as illustrated in Fig. \ref{fig:packet_count}, resulting in sharply increased computational and storage costs. This severely limits its practicality in large-scale network environments.
Therefore, how to design a lightweight and expressive traffic representation method without sacrificing the discriminative power of the model has become a core issue that needs to be broken through in the field of encrypted traffic analysis.

Another increasingly prominent challenge in ETA is the open-world problem. The core of this problem lies in the requirement that a model must not only accurately recognize known categories but also reject a vast amount of traffic belonging to unknown categories. Existing studies have mainly followed two directions:
1) Confidence-based methods, e.g., AppScanner \cite{taylor2016appscanner}, Rimmer et al. \cite{rimmer2017automated}, OpenMax \cite{yang2021deep,webb2019applying}, which set a global threshold based on the confidence distribution of classifier outputs to distinguish known and unknown samples;
2) Feature-distance–based methods, e.g., SHE-Net \cite{liu2021towards}, FlowPrint \cite{van2020flowprint}, kLND \cite{dahanayaka2023robust}, TrafficGPT \cite{ginige2024trafficgpt}, which determine category membership by measuring the distance between a test sample and known classes in the feature space.
Although these methods have achieved progress, they still rely on fixed global decision strategies, overlooking inherent differences in inter-class discriminability. This “one-size-fits-all” strategy causes the model to be overly conservative for easily distinguishable categories while overly aggressive for more complex ones, thereby weakening its generalization capability in open-world scenarios.

To address the above two major challenges, this paper proposes a lightweight encrypted traffic representation method named Up-Down Flow Sequence (UDFS), along with a prototype-based adaptive-threshold open-world traffic recognition mechanism. This approach significantly reduces computational overhead while effectively enhancing the model’s representation capability and open-world classification performance. The core contributions are as follows:

\begin{itemize}
    \item We propose the UDFS representation, which abstracts each flow within a trace as the sum of its upstream and downstream traffic, forming a highly compact yet informative two-dimensional sequence. While preserving key behavioral patterns, UDFS achieves over an order-of-magnitude reduction in feature length, effectively balancing representational capacity and computational cost.
    \item We design a prototype-based adaptive-threshold open-world traffic recognition mechanism, which learns dynamic prototype boundaries for each known category. It adaptively adjusts the loss function and decision threshold during both training and inference, thereby enhancing the model’s classification capability.
    \item Extensive experiments under various scenarios, including coarse- and fine-grained classification, concept drift, and open-world settings, demonstrate that the proposed method consistently outperforms existing approaches in classification performance. 
\end{itemize}

The remainder of this paper is organized as follows: Section \ref{bg} reviews related work on traffic representation and the open-world problem; Section \ref{motivation} presents the design motivation of UDFS; Section \ref{main} details the workflow and algorithmic design of the proposed system; Section \ref{experiment} describes the experimental setup and analyzes the results; finally, Section \ref{conclusion} concludes the paper and outlines directions for future research.

\section{Related Work}
\label{bg}
In this section, we provide a detailed description of traffic classification studies related to our work, including traffic representation methods and the open-world problem.

\subsection{Traffic Representation}
\paragraph{Statistical Features}
Statistical feature–based methods focus on enhancing model discriminability through careful feature engineering. Early work, such as K-Fingerprinting \cite{hayes2016k}, quantified feature importance and revealed the critical role of feature selection. Inspired by this, researchers began exploring the fusion of multi-source features to strengthen representation capability. For example, AppScanner \cite{taylor2016appscanner} combined raw packet length sequences with statistical features, while FlowPrint \cite{van2020flowprint} introduced destination-related cluster features for correlation analysis.
To achieve more comprehensive traffic characterization, subsequent studies have aimed to construct larger-scale, multi-dimensional feature sets. Shen et al. \cite{shen2019encrypted} designed hundreds of features spanning packet length, temporal patterns, and burst behaviors. Akbari et al. \cite{akbari2022traffic} and Guarino et al. \cite{guarino2022contextual} further integrated TLS handshake information, contextual data, and transport-layer payload features to achieve multi-perspective feature representation.
Although these methods improve performance to some extent through systematic feature combination, their heavy reliance on manual expertise and domain knowledge, as well as the complexity of feature engineering processes, make them difficult to adapt rapidly to the dynamic nature of encrypted traffic environments.

\paragraph{Byte Features}
To avoid reliance on labor-intensive manual feature engineering, Wang et al. \cite{wang2017end} were among the first to propose using raw traffic bytes directly as model input and performing traffic identification via end-to-end deep learning. Since then, substantial research has focused on improving model architectures and integrating multiple features. For example, methods such as PERT \cite{he2020pert}, ET-BERT \cite{lin2022bert}, PTU \cite{peng2024ptu}, Netmamba \cite{wang2024netmamba}, MIETT \cite{chen2025miett}, TrafficFormer \cite{zhou2025trafficformer}, ASNet \cite{peng2025bottom}, and TrafficLLM \cite{cui2025trafficllm} treat traffic byte sequences as sequences analogous to natural language and leverage pre-trained architectures such as BERT, Mamba, or large language models (LLMs) for feature learning.
Meanwhile, works such as YATC \cite{zhao2023yet}, ATVITSC \cite{liu2024atvitsc}, TFE-GNN \cite{zhang2023tfe}, SAT-Net \cite{li2025sat}, DE-GNN \cite{han2024gnn}, and MH-Net \cite{zhang2025revolutionizing} attempt to map byte sequences into images or graph structures to capture latent spatial dependencies and topological information.
Although these methods achieve very high accuracy on public datasets, Zhao et al. \cite{zhao2025sweet} pointed out that due to the pseudo-random nature of payloads, the reported high classification accuracy is often the result of “feature shortcuts” arising from dataset defects. In contrast, side-channel features in encrypted traffic—such as sequences composed of packet lengths and directions—are more reliable for recognition and generalization \cite{wickramasinghe2025sok}.

\paragraph{Sequence Features}
Existing sequence-based modeling methods can generally be divided into flow-level and trace-level approaches.
Flow-level methods, such as FS-Net \cite{liu2019fs}, GGFAST \cite{piet2023ggfast}, FlowPic \cite{shapira2019flowpic,shapira2021flowpic}, Chen et al. \cite{chen2020length,chen2022length}, Wu et al. \cite{wu2022online}, and METC-MVAE \cite{cai2022metc}, mainly focus on modeling side-channel sequence features (e.g., packet length, inter-arrival time, and direction) within individual flows. These methods leverage deep architectures such as CNNs, LSTMs, GRUs, and BERT to capture temporal dependencies and structural patterns. However, since a single flow often contains limited information and cannot fully reflect complex interaction behaviors, such methods struggle to discriminate fine-grained applications with highly similar traffic patterns.
In contrast, trace-level methods, e.g., Deep Fingerprinting \cite{sirinam2018deep}, RLTree \cite{li2021rltree}, TSHC-SW \cite{chen2023classify}, Exosphere \cite{fu2024detecting}, and Qu et al. \cite{qu2023input} treat multiple flows or entire traces as unified modeling units. It allows them to integrate richer contextual information and capture more fine-grained behavioral characteristics, achieving substantially higher classification accuracy than flow-level approaches. However, these methods must process extremely long and voluminous sequences, resulting in prohibitive computational and storage overheads, which severely limit their applicability in high-concurrency and large-scale network environments.

\subsection{Open-world Challenge}
To address the open-world challenge, existing studies have evolved along two main technical pathways. The first category of methods is based on model confidence. Representative approaches such as AppScanner \cite{taylor2016appscanner}, Rimmer et al. \cite{rimmer2017automated}, and OpenMax \cite{yang2021deep,webb2019applying} leverage the confidence scores produced by classifiers, using a predefined threshold to separate known samples from unknown ones.
The second category focuses on distance or similarity metrics in the feature space. Methods such as SHE-Net \cite{liu2021towards}, FlowPrint \cite{van2020flowprint}, kLND \cite{dahanayaka2023robust}, and TrafficGPT \cite{ginige2024trafficgpt} compute the similarity or distance between a test sample and the centroids of known classes, and similarly rely on a threshold for classification.
Although both approaches enhance recognition performance in open-world scenarios to some extent, their reliance on a unified and fixed decision threshold fails to accommodate the inherent differences in separability among classes, thus limiting the model’s classification capability in complex environments.

In summary, although extensive research has been conducted on encrypted traffic representation and open-world challenges, designing a traffic sequence representation that simultaneously achieves lightweight efficiency and strong discriminative capability, and further enhances the effectiveness of existing open-world handling strategies, remains a crucial and open research direction that demands deeper exploration.

\section{Design Motivation of UPFS}
\label{motivation}

This section introduces the motivation behind the proposed method, focusing on the underlying challenges of sequence-based features and the discovery of intrinsic patterns.

\begin{figure}[h]
\centering
\includegraphics[width=\linewidth]{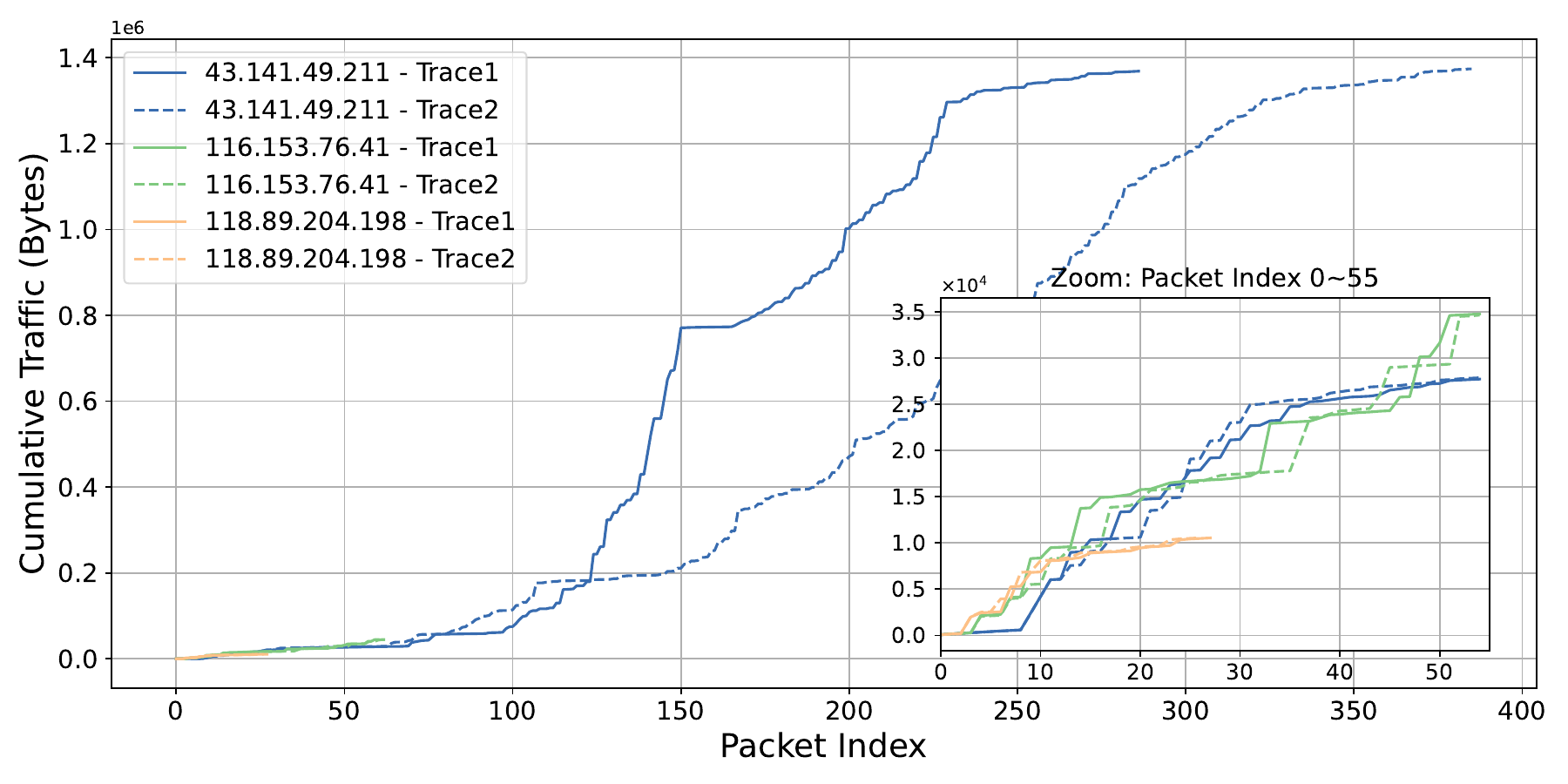}
\caption{Packet length sequence analysis of \texttt{Zhihu.com}. Three flows carrying identical resources across two traces are selected for comparison.}
\label{fig:single_flow_pkl}
\end{figure}

Beyond the trade-off between representational ability and computational cost, packet sequence–based methods face a more fundamental challenge of robustness. In real-world dynamic networks, factors such as network congestion, server load balancing, CDN scheduling dynamics, and the widespread adoption of multiplexing protocols like QUIC can cause the same application to exhibit drastically different packet length and timing sequences across sessions. This intrinsic instability of traffic patterns severely undermines model generalization, making it difficult to sustain high accuracy in complex and dynamic production environments.
As shown in Fig.~\ref{fig:single_flow_pkl}, two requests for the same network resource exhibit significant differences in their packet length sequences. This phenomenon undermines the stability and reproducibility of traditional sequence features, posing a direct threat to the generalization capability of models.

\begin{figure}[h]
\centering
\includegraphics[width=\linewidth]{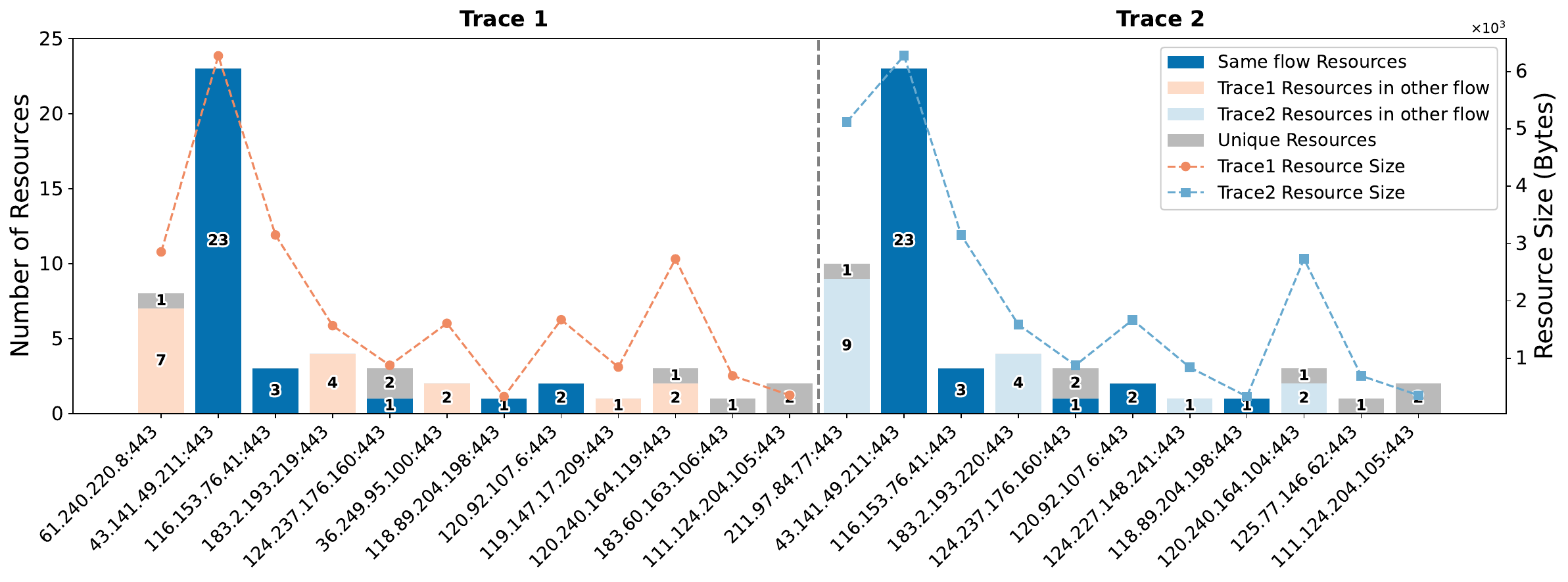}
\caption{Resource distribution and transmission volume of two traces of \texttt{Zhihu.com}.}
\label{fig:resources_flows}
\end{figure}

To explore more robust traffic features under network fluctuations, we conducted an in-depth decryption and content association analysis on large-scale encrypted traffic, aiming to reveal the intrinsic relationships between network flows and the application-layer resources they carry. Our study found that, although the underlying packet transmission patterns are highly unstable, there exist significant stability patterns at higher levels of abstraction. For example, through the analysis of access traffic to popular websites such as \texttt{Zhihu.com}, as shown in Fig.~\ref{fig:resources_flows}, we observed two key patterns:
\begin{enumerate}
\item \textbf{Stable Flow-Resource Mapping}: Although the responding server IP for resources may change across visits due to CDN scheduling (e.g., \texttt{183.2.193.219} vs. \texttt{183.2.193.220}), and the absolute order of flows may vary, the application-layer resource content carried by a single flow (e.g., a \texttt{loading.js}) is generally stable. More importantly, the total transmission volume of the flow corresponds directly to the resource size, remaining highly consistent across sessions. This suggests that flow-level transmission load is a more stable feature than packet length sequences.
\item \textbf{Localized Impact of Personalization}: Personalized content in modern webpages, such as user-tracking scripts and advertisements, can indeed cause variations in certain flows between sessions, e.g., traffic from \texttt{11.124.204.105}, in the Fig.~\ref{fig:resources_flows}. However, thanks to the templated design of webpages, the core resources that form the website’s main functionality and layout (e.g., CSS, JavaScript) remain unchanged. This means that the impact of personalization is localized, while most flows form a stable “baseline fingerprint.” It not only preserves the recognizability of traffic features but also implies that flows affected by personalization can themselves serve as differential signals for identifying personalized behaviors. 
\end{enumerate}

\begin{figure}[h]
    \centering
    \includegraphics[width=\linewidth]{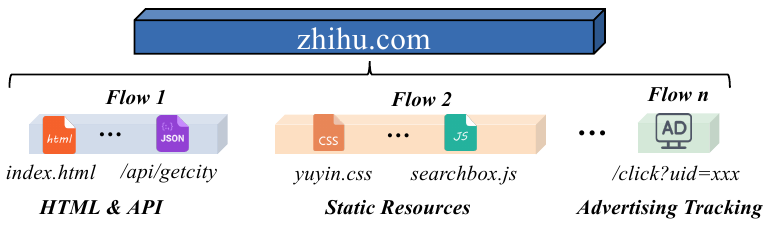}
    \caption{Mapping between actual resource loads and flows of \texttt{Zhihu.com}.}
    \label{fig:resources_lists}
\end{figure}

The above findings can ultimately be abstracted into the model illustrated in Fig. \ref{fig:resources_lists}, a complete user behavior (Trace) is composed of a sequence of flows, each transmitting specific resource combinations. A single flow is analogous to a word in a sentence, potentially carrying common resources; meanwhile, the temporal order and load composition of all flows within the entire trace resemble a complete sentence, forming a unique description of a specific webpage or application behavior.

This insight provides us with two key implications: \ding{182} The necessity of trace-level modeling: Compared to a single flow that may carry generic resources, such as jQuery.js, the combinatorial pattern formed by all flows (i.e., resources) within a trace possesses significantly greater discriminative power and is crucial for achieving fine-grained application identification. \ding{183} The direction for feature selection: Solely relying on raw and unstable packet sequences is insufficient for robust modeling. In contrast, flow-level aggregated statistical features that are directly related to resource size, particularly transmission volume, demonstrate notable robustness against network fluctuations. Therefore, modeling a trace as a sequence composed of these stable flow-level statistical features represents a promising approach to achieving both high discriminative power and robustness.

\section{Methodology}
\label{main}
\begin{figure*}[ht]
    \centering
    \includegraphics[width=\textwidth]{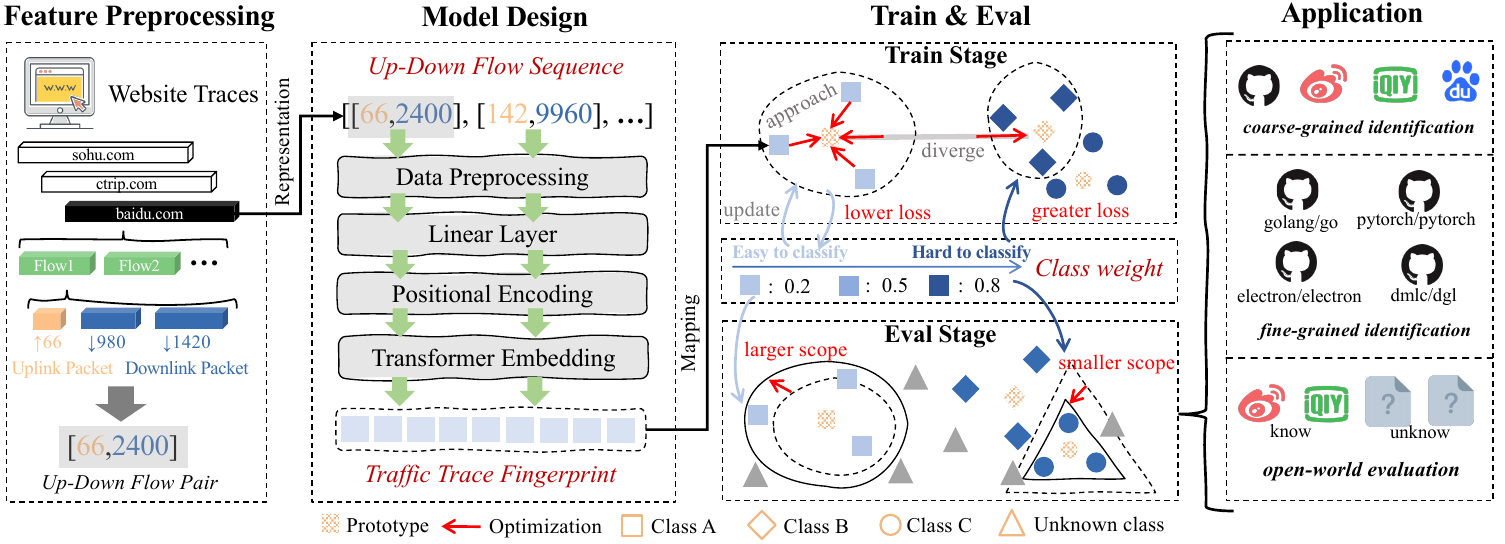}
    \caption{Overview of the proposed framework.}
    \label{fig:framework}
\end{figure*}

Based on the above findings, we first leverage the stability of flow-level transmission volume to design an efficient two-dimensional flow sequence representation method—UDFS. Building on this, we develop a Transformer-based deep feature extractor to capture traffic characteristics precisely. Finally, we propose an open-world recognition mechanism that integrates adaptive prototype learning with dynamic thresholding. The overall framework is illustrated in Fig. \ref{fig:framework}.

\subsection{UDFS Representation}

To address the instability of traditional packet-length sequence features in real-world networks, as well as the inherent trade-off between granularity and computational cost in flow-level and trace-level modeling, we propose the UDFS representation. UDFS abandons the highly volatile packet-level sequence and elevates the modeling unit to the flow level, characterizing each flow using more robust flow-level statistical features such as total transmitted volume. This design preserves essential behavioral information while effectively suppressing network-induced noise and significantly reducing computational complexity.

The core idea of UDFS lies in leveraging the inherent request–response pattern and its intrinsic asymmetry in network communications. In typical client–server interactions, uplink traffic primarily consists of resource requests—small in volume but structurally distinctive—whereas downlink traffic carries the actual content payloads, which are larger and more diverse. When these two directions are mixed, the substantial downlink traffic often overshadows the critical signaling patterns in the uplink. To mitigate this, UDFS explicitly separates bidirectional flows into uplink and downlink components, enabling the model to capture their distinct characteristics and provide a more fine-grained representation of application behaviors.

Based on this principle, we compress and represent a complete communication trace $T$, which consists of $N$ bidirectional flows $f_1, f_2, \dots, f_N$ arranged in chronological order, as a two-dimensional feature sequence $S_{UDFS}$. Specifically, for the $i$-th flow $f_i$ within the trace, we extract its total uplink bytes $B_{up}^{(i)}$ and total downlink bytes $B_{down}^{(i)}$, forming a two-dimensional vector $v_i = (B_{up}^{(i)}, B_{down}^{(i)})$. Consequently, the entire trace can be represented as:

\begin{eqnarray}
S_{UDFS} = [v_1, v_2, \dots, v_N], \quad  v_i \in \mathbb{R}^2
\end{eqnarray}

This representation achieves three key objectives. First, it drastically reduces the sequence length from tens of thousands of packets to merely dozens or hundreds of flows, fundamentally alleviating the computational and storage bottlenecks in trace-level analysis. Second, it preserves the macro-level temporal order of resource loading (i.e., the sequence of $v_i$). Finally, each vector $v_i$ encodes both the transmission load magnitude and the asymmetry between uplink and downlink traffic, providing a highly compact yet information-rich representation that significantly reduces the learning burden of subsequent models.

Moreover, considering the burstiness of network traffic and the wide dynamic range of flow sizes, directly using raw byte counts may lead to unstable model training. Therefore, we apply a logarithmic smoothing transformation to each flow value $b$ as follows:

\begin{eqnarray}
b' = \log(1+b)
\end{eqnarray}

This transformation effectively compresses the data scale and mitigates the impact of extreme values, enabling the model to focus on the underlying traffic variation patterns rather than absolute magnitudes.

\subsection{Transformer-based Feature Extractor}

The UDFS representation converts an encrypted traffic trace into a temporal feature sequence, where each element corresponds to a flow, and the ordering of elements reflects the temporal order of communication events. To learn deep semantic representations from this structured sequence—particularly to capture long-range dependencies that are crucial for application behavior recognition (e.g., the intrinsic relationship between the initial HTML document flow and subsequent CSS/JS or image resource flows)—we adopt a Transformer-based encoder as the core feature extraction module.

Our model architecture follows the standard Transformer Encoder design \cite{vaswani2017attention}, whose key components and processing flow are as follows:

\begin{enumerate}
\item \textbf{Input Embedding Layer}:  
The original two-dimensional flow vector $v_i \in \mathbb{R}^2$ in the UDFS sequence lies in a sparse feature space, making it difficult to perform complex nonlinear transformations directly. Therefore, we first apply a linear projection (i.e., an input embedding layer) to map each vector into a higher-dimensional, denser feature space $\mathbb{R}^{D_{model}}$. This step provides richer and more expressive representations for subsequent feature learning. Here, $D_{model}$ denotes the hidden dimension of the model and the working dimensionality of all subsequent layers. In our implementation, we set $D_{model}=128$.

\item \textbf{Positional Encoding}:  
The core of the Transformer architecture—the self-attention mechanism—cannot inherently perceive the order of input elements. However, in traffic analysis, the order in which flows occur carries crucial discriminative information. To address this, we introduce a deterministic sinusoidal positional encoding. The positional encoding vectors are added element-wise to the flow embeddings, injecting prior knowledge about each flow’s absolute position within the trace. This enables the model to distinguish and utilize the temporal order of communication behaviors.

\item \textbf{[CLS] Token for Global Aggregation}:  
Following the successful practice in BERT \cite{devlin2019bert}, we prepend a special learnable vector `[CLS]' to each UDFS sequence. This token acts as a global information aggregator that does not correspond to any specific flow but interacts with all flow representations through the self-attention mechanism. After several layers of Transformer encoding, the final output vector of the `[CLS]' token is regarded as a highly condensed global representation of the entire trace, which is directly used for downstream classification tasks.

\item \textbf{Transformer Encoder}:  
The main body of the model consists of $N_{layers}$ identical Transformer encoder layers. Each encoder layer contains two core submodules: a Multi-Head Self-Attention module and a Position-wise Feed-Forward Network, each followed by residual connections and layer normalization. The multi-head self-attention mechanism serves as the key to capturing global context, dynamically computing the dependency weights between any pair of flows within the sequence, allowing each flow representation to integrate global contextual information from the trace. The feed-forward network further performs nonlinear transformations on the attention outputs, enhancing the model’s expressive capacity. In our configuration, we set $N_{layers}=4$.
\end{enumerate}

In summary, when a UDFS sequence is fed into the model, it undergoes embedding, positional encoding, and multiple layers of hierarchical abstraction through $N_{layers}$ Transformer encoders, eventually producing a fixed-dimensional feature vector $h \in \mathbb{R}^{D_{model}}$. This vector $h$ serves as the trace fingerprint, encapsulating the key behavioral patterns and intrinsic structure of the encrypted traffic trace, and is directly used as the input to the downstream classifier for final identification.

\subsection{Adaptive Prototypical Learning}
To guide the model in learning a more discriminative embedding space, where intra-class samples are tightly clustered and inter-class samples are well separated, we design an adaptive prototypical learning strategy to replace the conventional cross-entropy loss. This strategy consists of two components: the prototypical network loss and the adaptive loss weighting mechanism.

\subsubsection{Prototypical Network Loss}
The core idea is rooted in metric learning. During training, for each mini-batch, we dynamically compute a temporary prototype $p_c$ for every class $c$ appearing in the batch. Each prototype is defined as the centroid of all feature vectors $h_i$ belonging to class $c$ within that batch:
$$p_c = \frac{1}{|S_c|} \sum_{h_i \in S_c} h_i,$$
where $S_c$ denotes the set of feature vectors corresponding to class $c$ in the current batch.  
Instead of relying on a fixed fully connected layer for classification, the decision is made by computing the Euclidean distance $\text{dist}(\cdot, \cdot)$ between each sample $h_i$ and all class prototypes. The negative distances are treated as classification logits, optimized via the standard Softmax cross-entropy.  
The prototypical loss $L_{proto}$ for a sample $h_i$ is defined as:
$$L_{proto}(h_i) = -\log \frac{\exp(-\text{dist}(h_i, p_{y_i}))}{\sum_{j} \exp(-\text{dist}(h_i, p_j))},$$
where $y_i$ is the ground-truth label of $h_i$.  
By minimizing this loss, the model is explicitly encouraged to pull samples closer to their corresponding class prototypes while pushing them away from other class prototypes. Consequently, a discriminative feature space with clear inter-class separability and compact intra-class cohesion is formed.

\subsubsection{Adaptive Loss Weighting}
In real-world traffic data, different classes often exhibit varying levels of separability. Some classes display diverse behavioral patterns, leading to more dispersed feature distributions, or even overlap significantly with other classes, making them ``hard'' categories for classification.  
To enable the model to dynamically focus on these challenging categories during training, we introduce an adaptive loss weighting mechanism. This mechanism periodically (e.g., after each epoch) computes a confusion score for every class, which reflects its separability difficulty, as shown in Algorithm~\ref{alg:weight_calculation}.  
Classes with higher confusion scores are assigned larger weights in the prototypical loss function, guiding the model to allocate more optimization capacity to refine the decision boundaries of these difficult categories.

\begin{algorithm}[h!]
\caption{Adaptive Weight Calculation}
\label{alg:weight_calculation}
\KwIn{Training set $D_{train}$, feature extractor $F(\cdot)$, number of neighbors $K$}
\KwOut{Class weights $\{W_c\}$}

\SetKwFunction{CalculateGlobalPrototypes}{CalculateGlobalPrototypes}
\SetKwProg{Fn}{Function}{:}{}

\Fn{\CalculateGlobalPrototypes{$D_{train}, F$}}{
    $H \leftarrow \{\}$ \tcp*[f]{Initialize feature lists for each class}\\
    \ForEach{$(x_i, y_i) \in D_{train}$}{
        $h_i \leftarrow F(x_i)$ \\
        Append $h_i$ to $H[y_i]$\;
    }
    $\mathcal{P} \leftarrow \{\}$ \tcp*[f]{Initialize global prototype dictionary}\\
    \ForEach{class $c \in H$}{
        $\mathcal{P}_c \leftarrow \text{mean}(H[c])$\;
    }
    \Return{$\mathcal{P}, H$}
}

$\mathcal{P}, H \leftarrow$ \CalculateGlobalPrototypes{$D_{train}, F$} \tcp*[f]{Compute all global prototypes and features}\\
$W \leftarrow \{\}$ \tcp*[f]{Initialize weight dictionary}\\
\ForEach{class $c \in \mathcal{P}$}{
    $D_{intra}(c) \leftarrow \text{mean}(\{\text{dist}(h, \mathcal{P}_c) \mid h \in H[c]\})$ \tcp*[f]{Intra-class average distance}\\
    $D_{inter}(c) \leftarrow \text{mean}(\text{TopK-Nearest}(\{\text{dist}(\mathcal{P}_c, \mathcal{P}_j) \mid j \neq c\}, K))$ \tcp*[f]{Inter-class nearest distances}\\
    $W_c \leftarrow D_{intra}(c) / (D_{inter}(c) + \epsilon)$\;
}
\Return{$W$}
\end{algorithm}

\subsection{Open-World Recognition with Dynamic Thresholding}

Traditional closed-set classifiers assume that all test samples belong to the set of known categories seen during training. To endow the model with the capability to recognize “unknown” traffic, we design an open-set recognition mechanism based on dynamic thresholding. This mechanism first computes a dedicated decision boundary $\theta_c$ for each category after training, based on its distribution characteristics and separability difficulty, as shown in Algorithm \ref{alg:threshold_calculation}. During inference, this boundary is used to judge new samples, as detailed in Algorithm \ref{alg:inference}.

\begin{algorithm}[h!]
\caption{Dynamic Threshold Calculation}
\label{alg:threshold_calculation}
\KwIn{Training set $D_{train}$, feature extractor $F(\cdot)$, percentile $p$, adjustment hyperparameter $\alpha$}
\KwOut{Dynamic threshold set $\{\theta_c\}$ for each class}

$\mathcal{P}, H \leftarrow$ \text{CalculateGlobalPrototypes}($D_{train}, F$) \tcp*[f]{Call function from Algorithm 1}\\
$\theta^{base}, S \leftarrow \{\}, \{\}$ \tcp*[f]{Initialize base thresholds and difficulty scores}\\

\ForEach{class $c \in \mathcal{P}$}{
    $Dists_c \leftarrow \{\text{dist}(h, \mathcal{P}_c) \mid h \in H[c]\}$ \;
    $\theta_c^{base} \leftarrow \text{Percentile}(Dists_c, p)$ \;
    $D_{intra}(c) \leftarrow \text{mean}(Dists_c)$ \;
    $D_{inter}(c) \leftarrow \text{mean}(\text{TopK-Nearest}(\{\text{dist}(\mathcal{P}_c, \mathcal{P}_j) \mid j \neq c\}, K))$ \;
    $S_c \leftarrow D_{intra}(c) / (D_{inter}(c) + \epsilon)$ \;
}

$Z_S \leftarrow \text{Z-Score-Normalize}(\{S_c\})$ \tcp*[f]{Standardize difficulty scores}\\
$\theta \leftarrow \{\}$ \;

\ForEach{class $c \in S$}{
    $\gamma_c \leftarrow 1.0 - \alpha \cdot \tanh(Z_S[c])$ \;
    $\theta_c \leftarrow \theta_c^{base} \cdot \gamma_c$ \;
}

\Return{$\theta$}
\end{algorithm}

\begin{algorithm}[h!]
\caption{Open-World Inference}
\label{alg:inference}
\KwIn{Test sample feature $h_{test}$, global prototype set $\{\mathcal{P}_c\}$, dynamic threshold set $\{\theta_c\}$}
\KwOut{Predicted label $\hat{y} \in \{C_1, \dots, C_N, \text{Unknown}\}$}

$d_{min} \leftarrow \infty$\;
$c^* \leftarrow \text{null}$\;

\ForEach{class $c \in \{\mathcal{P}_c\}$}{
    $d \leftarrow \text{dist}(h_{test}, \mathcal{P}_c)$\;
    \If{$d < d_{min}$}{
        $d_{min} \leftarrow d$\;
        $c^* \leftarrow c$\;
    }
}

\If{$d_{min} \le \theta_{c^*}$}{
    $\hat{y} \leftarrow c^*$\;
}
\Else{
    $\hat{y} \leftarrow \text{Unknown}$\;
}

\Return{$\hat{y}$}
\end{algorithm}

This mechanism constructs a customized dynamic decision boundary for each known category, taking into account both its intra-class distribution and inter-class relationships. As a result, the model is not only able to accurately classify known applications but also effectively reject any samples that do not conform to known patterns, achieving robust unknown traffic recognition.

\section{Experiments}
\label{experiment}

To comprehensively evaluate the performance of our proposed encrypted traffic recognition framework, we design a series of experiments aimed at answering the following three core questions:
\begin{enumerate}
\item \textbf{Q1 Overall Performance Evaluation:} How does our method perform compared to existing baselines in various complex scenarios, such as different granularity of encrypted traffic classification tasks, concept drift, and open-world recognition?
\item \textbf{Q2 Ablation Study:} How effective are the proposed UPFS feature representation and open-world traffic recognition mechanisms?
\item \textbf{Q3 Robust Analysis:} How robust and generalizable is the proposed framework across different hyperparameter settings and feature extractor architectures?
\end{enumerate}

\subsection{Experimental Setup}

\subsubsection{Datasets and Evaluation Scenarios}
Existing public encrypted traffic datasets generally suffer from outdated protocols, limited scenarios, and difficulty in capturing complete single-user access traces \cite{wickramasinghe2025sok}. To rigorously evaluate model performance under conditions closer to modern network environments, we construct and annotate six high-quality datasets totaling approximately 119.5GB (the code and data will be publicly released at \url{https://github.com/kid1999/UDFS}). In this work, we focus on two representative datasets:

\begin{itemize}
\item \textbf{GitHub}: A \textbf{fine-grained} classification dataset, containing traffic generated from homepage visits to 10 popular code repositories with highly similar functions and interaction patterns (e.g., `vuejs/vue'). This dataset aims to challenge the model’s ability to distinguish subtle behavioral differences.
\item \textbf{Baidu}: A \textbf{coarse-grained} classification dataset, containing traffic from homepage visits to 20 well-known websites with distinct traffic patterns and page structures (e.g., `baidu.com', `weibo.com', `qq.com', etc.).
\end{itemize}

To comprehensively assess the robustness and generalization ability of the model in realistic dynamic environments, we design the following two challenging evaluation scenarios:

\begin{enumerate}
\item \textbf{Concept Drift}: This scenario simulates changes in traffic patterns over time caused by application updates, content changes, or network environment variations. We collect traffic from the same set of applications at different time periods (e.g., one month apart), train the model on earlier data, and test it on later data to evaluate temporal robustness.
\item \textbf{Open-World Recognition}: This scenario evaluates the model’s ability to recognize and reject unknown applications. We train the model on $N$ known classes, but the test set contains these $N$ known classes mixed with a much larger number of unknown classes. For example, in the \textbf{Baidu} test set, approximately \textbf{10,000} unknown websites are mixed in; in the \textbf{GitHub} test set, traffic from 90 unknown repositories is included.
\end{enumerate}

\subsubsection{Baseline Models, Evaluation Metrics, and Implementation Details}
To validate the effectiveness of our proposed method, we select nine representative baseline models covering current mainstream approaches for comprehensive comparison:

\begin{itemize}
\item \textbf{Statistical Features}: Models using statistical or IP-based features extracted from flows, e.g., AppScanner \cite{taylor2016appscanner} (statistical features) and FlowPrint \cite{van2020flowprint} (IP features).
\item \textbf{Byte Features}: Models directly using raw byte sequences as input, including ET-BERT \cite{lin2022bert} and TrafficFormer \cite{zhou2025trafficformer}.
\item \textbf{Sequence Features}: Methods that use statistical or sequential features of individual flows as analysis units, e.g., FS-Net \cite{liu2019fs} (deep learning-based), and methods that use multiple flows or complete traces as units, including Deep Fingerprinting (DF) \cite{sirinam2018deep} and TraceClassifier \cite{qu2023input}.
\item \textbf{Open-World Evaluation Methods}: OpenMax \cite{yang2021deep,webb2019applying} and AppScanner \cite{taylor2016appscanner} based on model confidence distributions, as well as kLND \cite{dahanayaka2023robust,ginige2024trafficgpt} and FlowPrint \cite{van2020flowprint} based on distance similarity.
\end{itemize}

For performance evaluation, we adopt standard classification metrics, including Accuracy (ACC), Precision, Recall, and F1-Score. To fairly evaluate model performance on imbalanced datasets, all multi-class metrics are computed using Macro-Averaging.

All experiments are conducted using the PyTorch deep learning framework on a server configured with dual Intel Xeon Gold 6330 CPUs and a single NVIDIA A100 (80 GB) GPU. During training, we adopt the AdamW optimizer with a learning rate of $1 \times 10^{-4}$. The batch size is set to 64. All baseline models are trained and evaluated on our datasets using their default hyperparameter settings as recommended in their source code to ensure fair comparison.

\begin{table*}[htbp]
\centering
\caption{Comparison of closed-set performance between the proposed method and baseline models on the GitHub and Baidu datasets, including both standard and concept drift scenarios.
}
\label{tab:closed_set}
\renewcommand{\arraystretch}{1.2}
\resizebox{\linewidth}{!}{ %
\begin{tabular}{l|cccc|cccc|cccc|cccc}
\toprule
\multirow{2}{*}{\textbf{Method}} & \multicolumn{4}{c|}{\textbf{Github}} & \multicolumn{4}{c|}{\textbf{Github (Concept Drift)}} & \multicolumn{4}{c|}{\textbf{Baidu}} & \multicolumn{4}{c}{\textbf{Baidu (Concept Drift)}} \\
\cmidrule(lr){2-5} \cmidrule(lr){6-9} \cmidrule(lr){10-13} \cmidrule(lr){14-17}
 & ACC & Precision & Recall & F1 & ACC & Precision & Recall & F1 & ACC & Precision & Recall & F1 & ACC & Precision & Recall & F1 \\
\midrule
AppScanner   & 32.44 & 34.57 & 32.19 & 32.43 & 24.66 & 44.39 & 22.54 & 23.78 & 69.98 & 78.26 & 70.71 & 72.34 & 63.66 & 71.71 & 63.96 & 65.07 \\
FlowPrint   & 30.07 & 11.45 & 27.66 & 15.70 & 6.15  & 1.68  & 6.37  & 2.64  & 95.12 & 91.12 & 94.59 & 91.16 & 97.62 & 97.37 & 96.75 & 96.98 \\  \midrule
FS-Net      & 16.01 & 20.37 & 14.80 & 13.66 & 15.20 & 12.30 & 14.23 & 11.67 & 92.26 & 93.14 & 91.56 & 92.17 & 90.88 & 91.81 & 89.83 & 90.52 \\
DF          & 66.00 & 74.00 & 66.00 & 65.32 & 39.30 & 53.79 & 39.30 & 35.85 & 97.96 & 98.33 & 98.00 & \underline{98.01} & 97.10 & 97.29 & 97.10 & 97.07 \\
TraceClassifier & 96.00 & 96.42 & 96.00 & 95.98 & 76.37 & 81.25 & 76.69 & 75.48 & 100.00 & 100.00 & 100.00 & \color{red}\textbf{100.00} & 98.00 & 98.07 & 98.00 & 97.97 \\ \midrule
ET-BERT     & 99.77 & 99.73 & 99.73 & \color{red}\textbf{99.73} & 9.19  & 9.59  & 9.40  & 8.58  & 96.47 & 95.14 & 95.29 & 95.20 & 48.15 & 44.66 & 44.56 & 42.70 \\
TrafficFormer & 11.43 & 1.14  & 10.00 & 2.05  & 14.31 & 1.43  & 10.00 & 2.50  & 80.49 & 85.85 & 79.26 & 80.40 & 71.09 & 74.00 & 67.18 & 68.27 \\
\midrule
w/o Prototype & 88.00 & 83.85 & 88.00 & 84.92 & 88.57 & 84.16 & 88.30 & 85.52 & 100.00 & 100.00 & 100.00 & \color{red}\textbf{100.00} & 99.85 & 99.85 & 99.85 & \color{red}\textbf{99.85} \\
w/o Weight    & 98.00 & 98.33 & 98.00 & \underline{98.04} & 97.56 & 97.97 & 97.50 & \underline{97.62} & 100.00 & 100.00 & 100.00 & \color{red}\textbf{100.00} & 99.85 & 99.85 & 99.85 & \color{red}\textbf{99.85} \\
w/o UPFS      & 98.00 & 98.18 & 98.00 & 97.99 & 23.83 & 29.10 & 24.38 & 20.33 & 100.00 & 100.00 & 100.00 & \color{red}\textbf{100.00} & 99.00 & 99.05 & 99.00 & \underline{99.00} \\
\rowcolor{gray!20}
Ours        & 98.00 & 98.18 & 98.00 & 97.99 & 97.56 & 97.65 & 97.54 & \color{red}\textbf{97.63} & 100.00 & 100.00 & 100.00 & \color{red}\textbf{100.00} & 99.85 & 99.85 & 99.85 & \color{red}\textbf{99.85} \\
\bottomrule
\end{tabular}
}
\end{table*}

\subsection{Overall Performance Evaluation (Q1)}

To comprehensively address \textbf{Q1}, we evaluate the performance of various methods across classification tasks of different granularity under closed-set, concept drift, and open-world scenarios. The experimental results are shown in Table \ref{tab:closed_set} and Table \ref{tab:open_set}.

\paragraph{Closed-Set Scenario Analysis.}
In standard closed-set classification tasks, deep learning methods significantly outperform traditional traffic fingerprinting and statistical feature-based methods. Taking the GitHub dataset as an example, TraceClassifier, ET-BERT, and our method all achieve over 95\% accuracy and F1 scores, while traditional methods such as FlowPrint and AppScanner fall below 30\%. Even in the relatively simpler Baidu dataset, traditional methods improve somewhat but still generally lag behind deep learning models. Overall, deep learning–driven approaches demonstrate stronger feature representation capability and generalization performance across different granularities of encrypted traffic classification, with clear advantages in accuracy and stability.

\paragraph{Concept Drift Scenario Analysis.}
The concept drift scenario simulates the evolution of traffic patterns caused by application updates or content changes in the real world, posing a severe challenge to model generalization and robustness. As shown in Table \ref{tab:closed_set}, most baseline models exhibit significant performance degradation under this scenario, especially in the fine-grained GitHub dataset.

Specifically, models relying directly on raw byte sequences show notable weaknesses in generalization under concept drift. For example, ET-BERT’s F1 scores drop sharply from 99.73\% and 95.20\% to 8.58\% and 42.70\% on the GitHub and Baidu datasets, respectively. This suggests that such models may rely on feature shortcuts rather than truly adapting to changes in traffic distribution, a conclusion consistent with findings from Wickramasinghe et al. \cite{wickramasinghe2025sok} and Zhao et al. \cite{zhao2025sweet}.

For models relying on specific metadata features, performance strongly depends on task granularity. FlowPrint, which depends on IP clustering, maintains a high F1 of 96.98\% on the coarse-grained Baidu dataset due to relatively stable IPs. However, in the fine-grained GitHub scenario, IP distinctions among different repositories are insufficient, and concept drift reduces F1 to 2.64\%, rendering it almost ineffective. Similarly, FS-Net also shows significant degradation in fine-grained tasks, achieving only 11.67\% F1, as expected — using a single flow as input tends to cause feature overlaps that hinder fine-grained differentiation.

In contrast, statistical feature–based and trace-based methods show relatively better robustness. TraceClassifier achieves F1 scores of 75.48\% and 97.97\% under concept drift in GitHub and Baidu, respectively, demonstrating the effectiveness of using complete access traces as the analysis unit, though some performance loss still occurs in fine-grained scenarios.

Our proposed method shows outstanding stability and generalization in both concept drift scenarios. In the most challenging GitHub (Concept Drift) scenario, our method achieves an F1 score of \textbf{97.63\%}, far exceeding all baselines. In the Baidu (Concept Drift) scenario, it achieves \textbf{99.85\%} F1, almost unaffected. These results demonstrate that our method learns more intrinsic traffic behavior patterns that are less sensitive to temporal changes, thereby possessing stronger temporal robustness.

\paragraph{Open-World Scenario Analysis.}
Open-world recognition aims to evaluate the model’s ability to accurately identify and reject unknown traffic, which is critical for ensuring security and reliability in real-world networks. As shown in Table \ref{tab:open_set}, we conducted rigorous testing of all models under this scenario, particularly in the fine-grained GitHub dataset, where unknown and known classes exhibit highly similar behavior patterns, posing a great challenge.

The results indicate that traditional encrypted traffic fingerprint methods struggle in open-world settings. For example, FlowPrint and AppScanner nearly fail in the fine-grained GitHub dataset, with F1 scores of only 0.25\% and 2.77\%, exposing the limitations of relying on IP or statistical features when distinguishing similar unknown applications. For a fairer comparison, we evaluated models under the same UPFS feature condition against mainstream open-world methods such as K-LND and OpenMax. On the relatively simple coarse-grained Baidu dataset, K-LND achieved a competitive F1 score of 84.13\%. However, in the more challenging fine-grained GitHub dataset, all baseline open-world methods suffered significant performance drops, with F1 scores failing to exceed 45\%, highlighting the difficulty of distinguishing known from unknown classes when inter-class differences are small.

In contrast, our proposed method achieves the best performance across both datasets. On the Baidu dataset, it attains an F1 score of \textbf{85.64\%}. Particularly in the extremely challenging GitHub dataset, our method achieves an F1 score of \textbf{54.85\%}, significantly outperforming all baseline models. This strongly validates the effectiveness and stability of our approach in addressing open-world recognition challenges.

\begin{table}[htbp]
\centering
\caption{Performance comparison of open-world settings.}
\label{tab:open_set}
\resizebox{\linewidth}{!}{ %
\begin{tabular}{l|cccc|cccc}
\toprule
\multirow{2}{*}{\textbf{Method}} & \multicolumn{4}{c|}{\textbf{Baidu Dataset (Open-World)}} & \multicolumn{4}{c}{\textbf{GitHub Dataset (Open-World)}} \\
\cmidrule(lr){2-5} \cmidrule(lr){6-9}
& ACC & Precision & Recall & F1 & ACC & Precision & Recall & F1 \\
\midrule
FlowPrint   & 17.47 & 30.50 & 92.71 & 41.73 & 1.06 & 0.23 & 9.13 & 0.25 \\
AppScanner  & 81.17 & 91.34 & 45.48 & 58.31 & 0.19 & 43.00 & 1.55 & 2.77 \\
\midrule
K-LND1   & 95.78 & 92.35 & 78.20 & \underline{84.13} & 58.90 & 35.78 & 78.32 & 42.36 \\
K-LND2   & 16.52 & 38.67 & 92.27 & 48.26 & 9.51  & 14.70 & 83.94 & 19.29 \\
K-LND3   & 95.72 & 91.90 & 78.34 & 84.03 & 58.08 & 35.80 & 78.42 & 42.33 \\
OpenMax  & 94.59 & 94.20 & 70.15 & 78.82 & 58.51 & 39.13 & 75.49 & 44.69 \\ \midrule
w/o Weighted Loss & 92.91 & 92.04 & 81.56 & 83.41 & 72.69 & 43.42 & 70.90 & \underline{45.28} \\
w/o Weighted Threshold   & 90.80 & 83.73 & 85.99 & 82.71 & 62.58 & 36.55 & 89.66 & 44.00 \\
w/o UPFS  & 91.31 & 84.12 & 86.06 & 83.03 & 23.73 & 29.10 & 24.28 & 20.23 \\
\rowcolor{gray!20} \textbf{Ours} & 95.26 & 88.09 & 85.12 & \color{red}\textbf{85.64} & 75.46 & 44.94 & 92.67 & \color{red}\textbf{54.85} \\
\bottomrule
\end{tabular}
}
\end{table}

\begin{figure*}[htbp]
\centering
\subfloat[\small Github Concept-Drift]{
\includegraphics[width=0.23\linewidth]{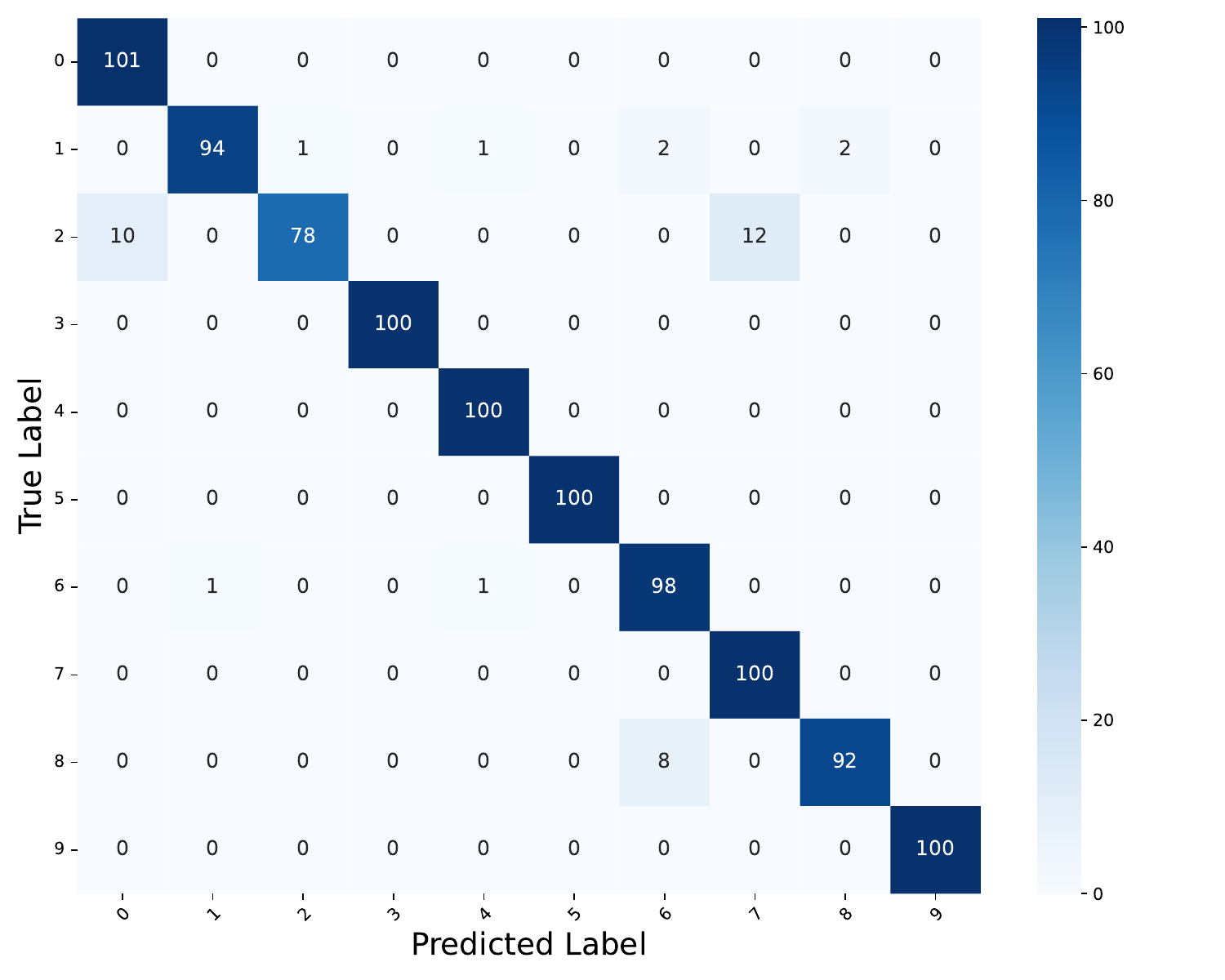}
    \label{subfig:github_confusion_matrix_closed_set}
    }
\hfill
\subfloat[\small Github Open-World]{
\includegraphics[width=0.23\linewidth]{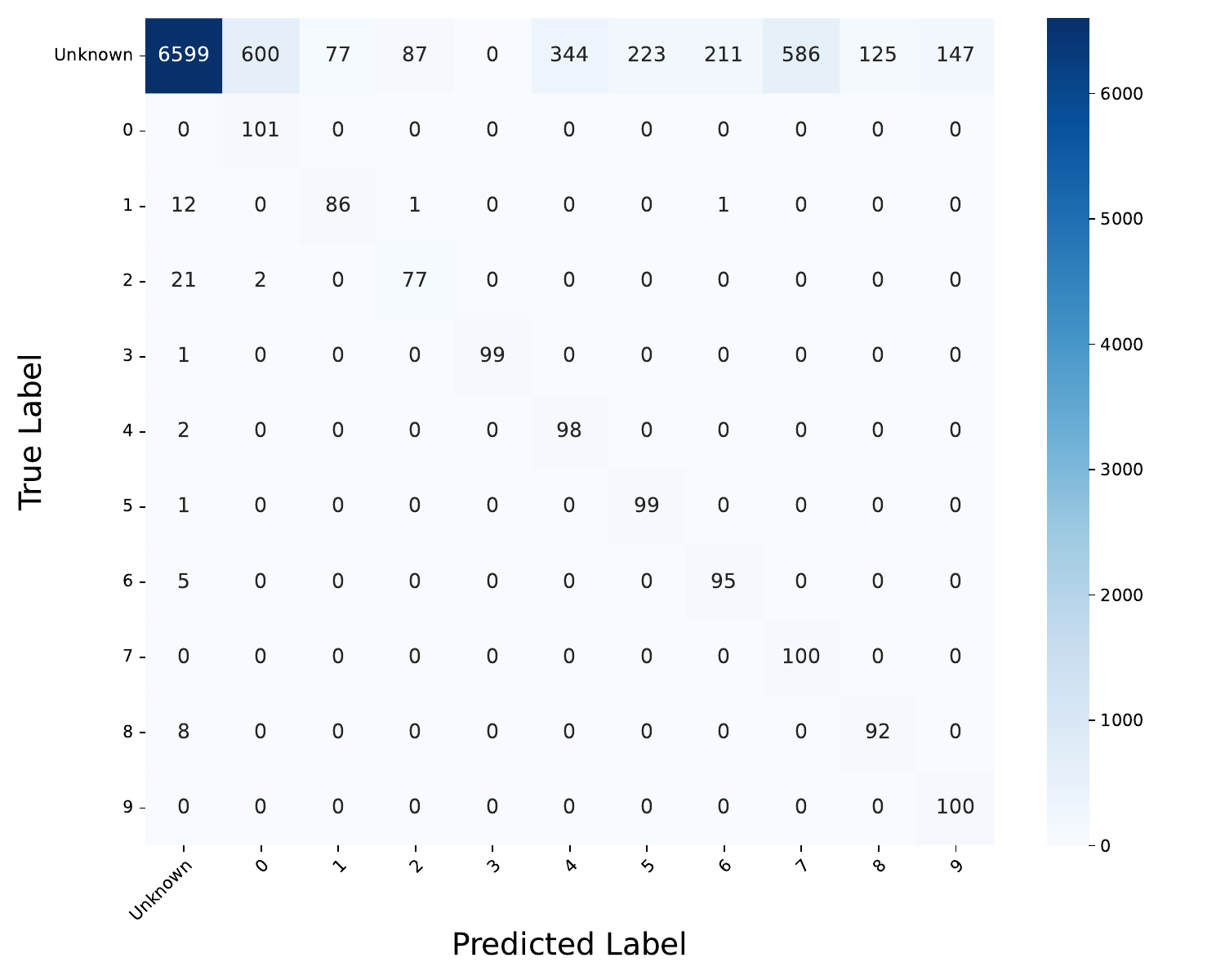}
    \label{subfig:github_confusion_matrix_open_set}
    }
\hfill
\subfloat[\small Baidu Concept-Drift]{
\includegraphics[width=0.23\linewidth]{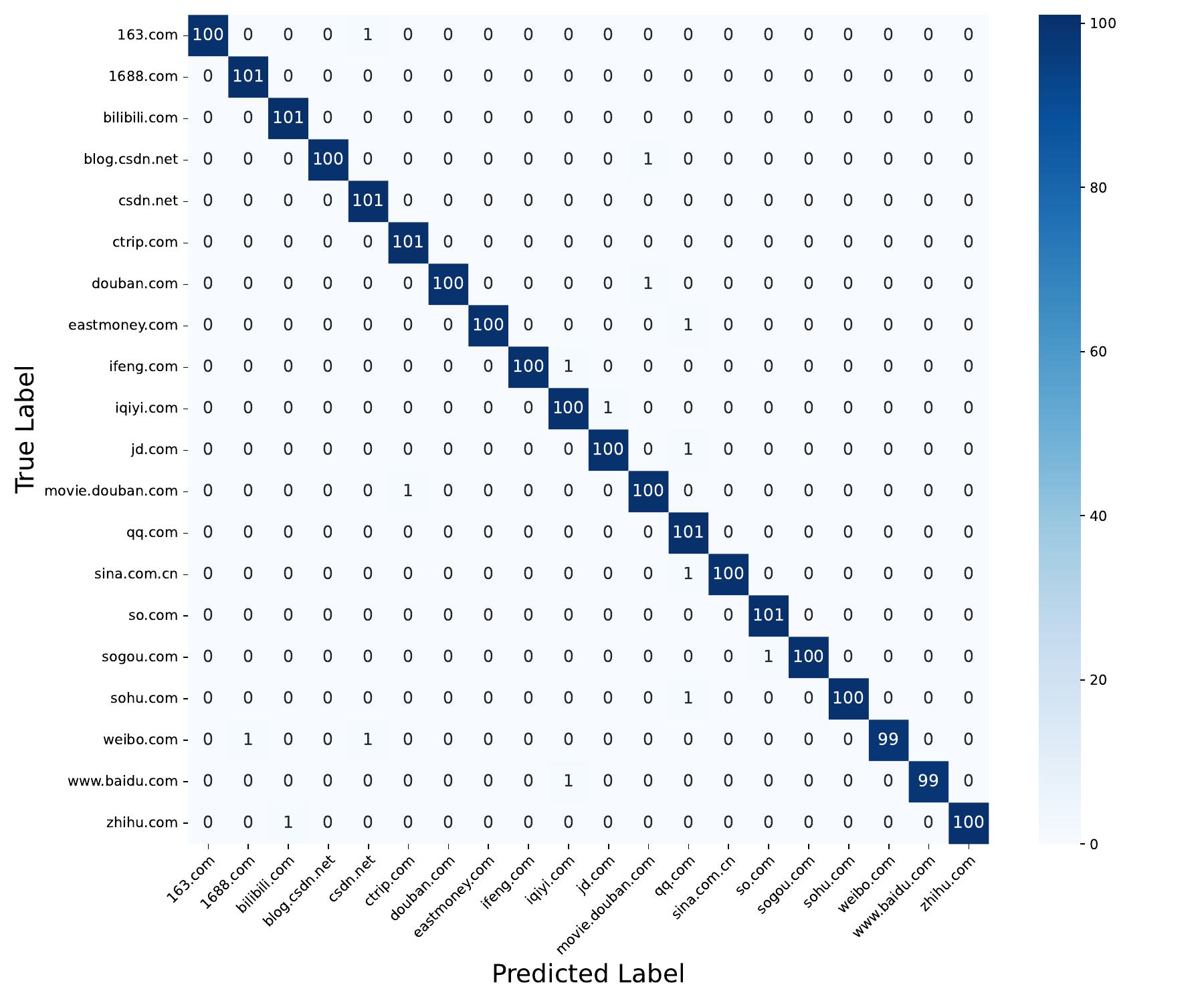}
    \label{subfig:baidu_confusion_matrix_closed_set}
}
\hfill
\subfloat[\small Baidu Open-World]{
\includegraphics[width=0.23\linewidth]{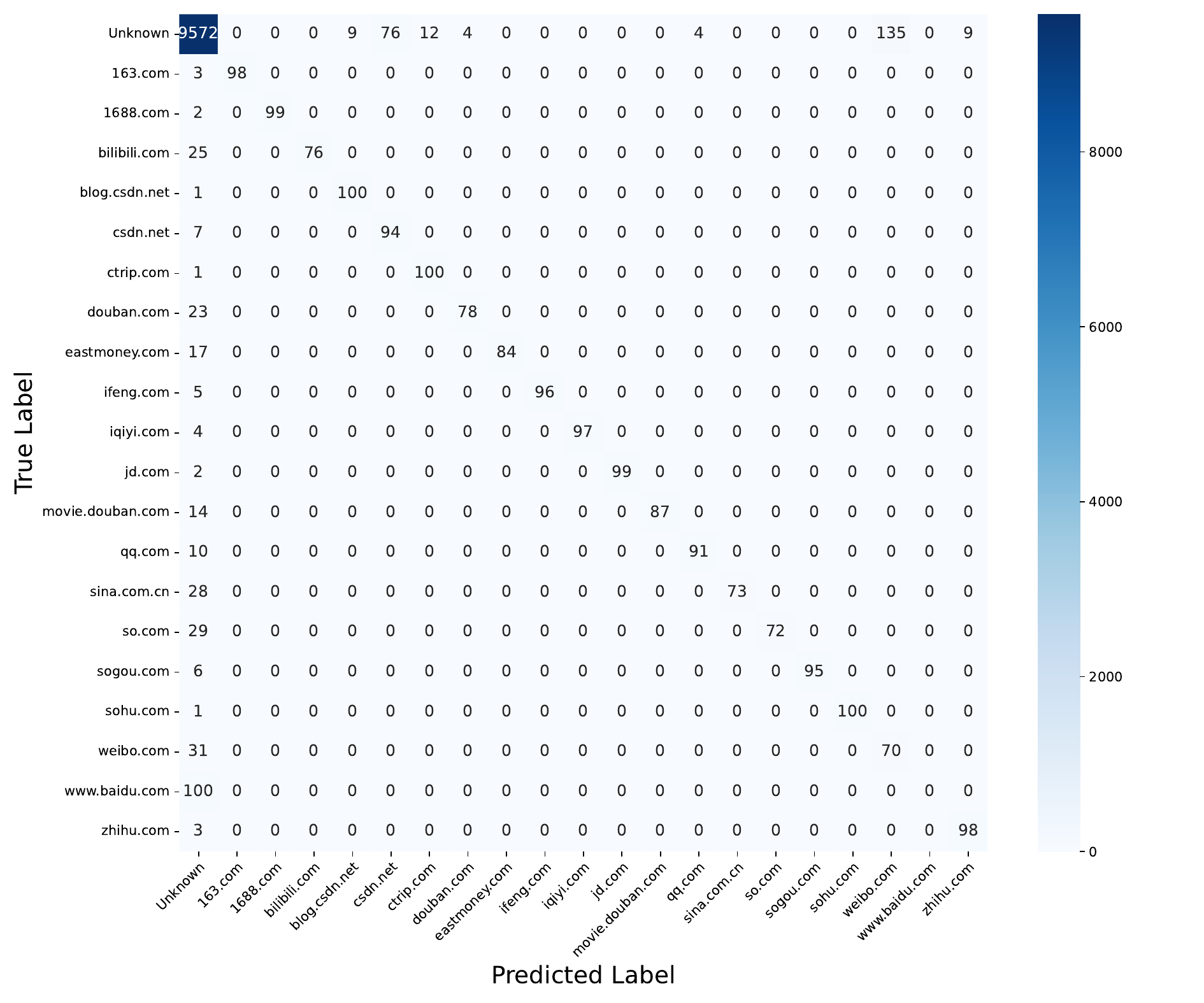}
    \label{subfig:baidu_confusion_matrix_open_set}
    }
\caption{Confusion matrices of the proposed method under concept drift and open-world scenarios.}
\label{fig:matrix}
\end{figure*}

To intuitively analyze the model’s fine-grained classification performance under complex scenarios, we present the confusion matrices for concept drift and open-world settings in Fig. \ref{fig:matrix}.  
In the concept drift scenario, the matrix exhibits a clear and concentrated diagonal, indicating that the model can robustly recognize most classes even when distributions change over time, demonstrating strong temporal robustness.  
In the open-world evaluation, the fine-grained nature of the GitHub dataset makes recognition significantly more challenging than the Baidu dataset, as unknown classes share highly similar traffic features with known classes. Nevertheless, our model still demonstrates the best performance on this dataset, validating the effectiveness of our approach. Naturally, we also observe that there remains room for improvement when handling highly similar traffic.  

\begin{mybox} 
\textbf{Answer for RQ1:} Our method not only achieves top-tier performance in standard closed-set classification tasks but, more importantly, demonstrates robustness and generalization far surpassing existing baselines under realistic challenges such as concept drift and open-world scenarios, showing strong potential for real-world application.
\end{mybox}

\subsection{Ablation Study (Q2)}

To verify the effectiveness of the key modules in our proposed framework, we conducted a series of ablation studies, focusing on two core components: the UPFS feature representation method and the open-world recognition mechanism.

\begin{figure*}[t!]
\centering
\subfloat[Packet Length Sequence]{
    \includegraphics[width=0.31\linewidth]{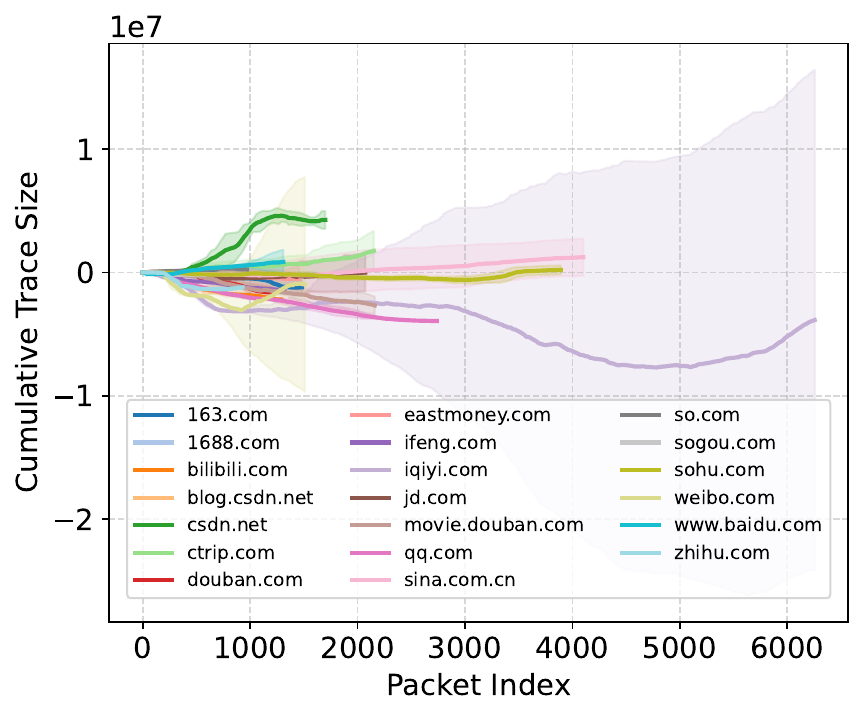}
    \label{subfig:baidu_feature_trace}
}
\hfill
\subfloat[Upstream Sequence]{
    \includegraphics[width=0.31\linewidth]{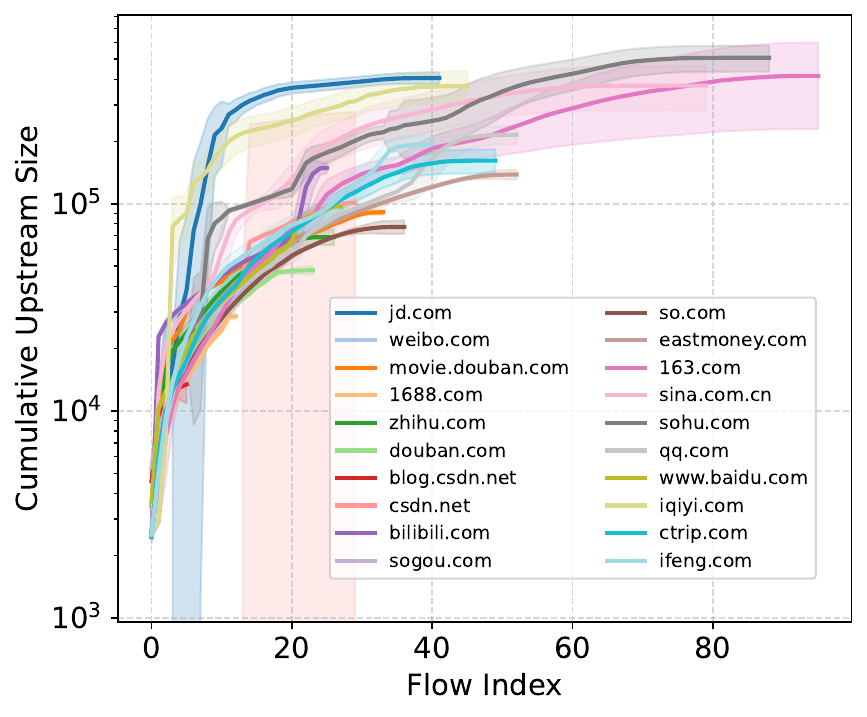}
    \label{subfig:baidu_feature_up}
}
\hfill
\subfloat[Downstream Sequence]{
    \includegraphics[width=0.31\linewidth]{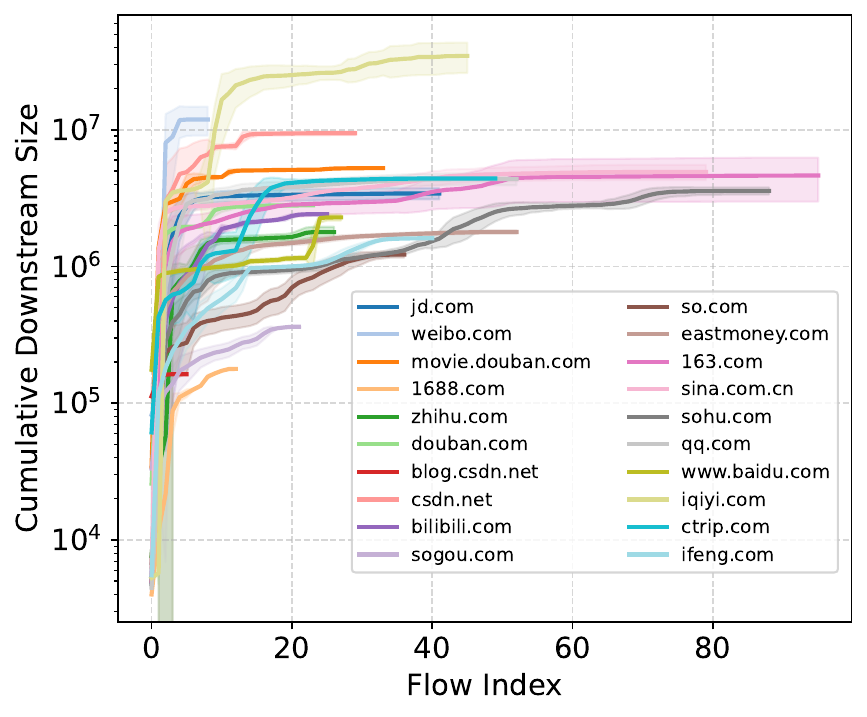}
    \label{subfig:baidu_feature_down}
}
\hfill
\subfloat[Packet Length Sequence]{
    \includegraphics[width=0.31\linewidth]{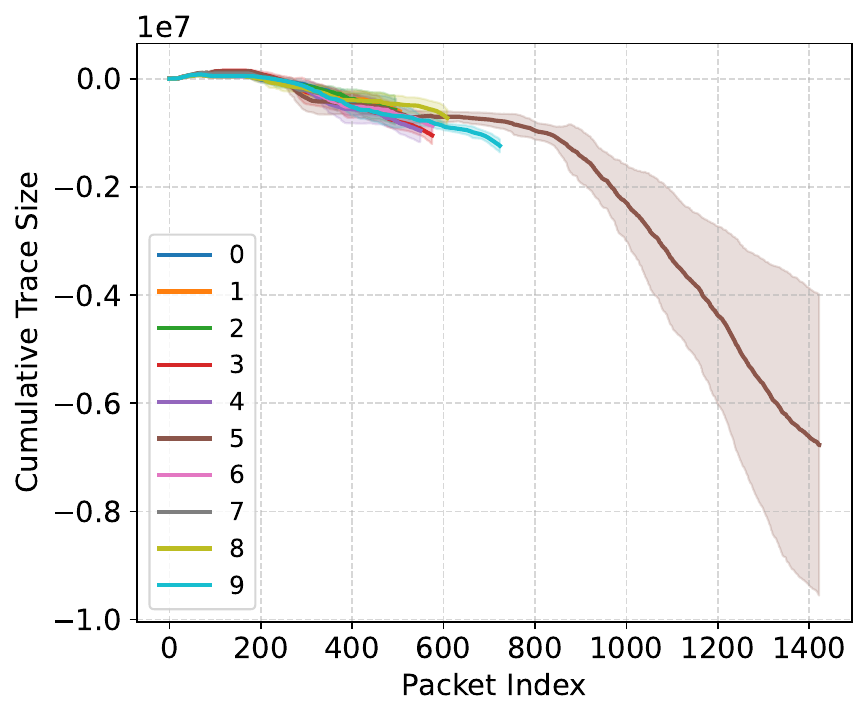}
    \label{subfig:github_feature_trace}
}
\hfill
\subfloat[Upstream Sequence]{
    \includegraphics[width=0.31\linewidth]{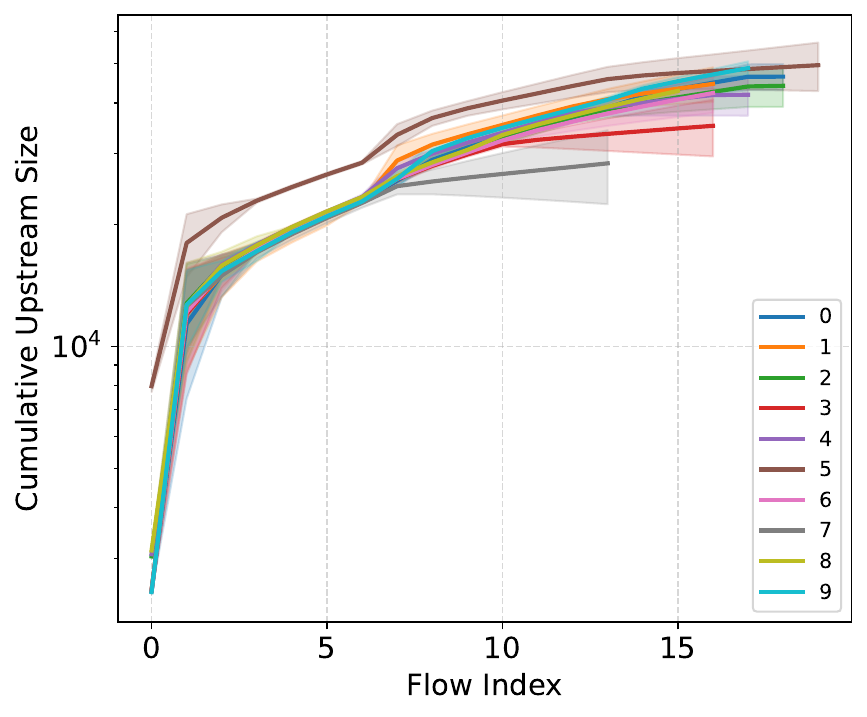}
    \label{subfig:github_feature_up}
}
\hfill
\subfloat[Downstream Sequence]{
    \includegraphics[width=0.31\linewidth]{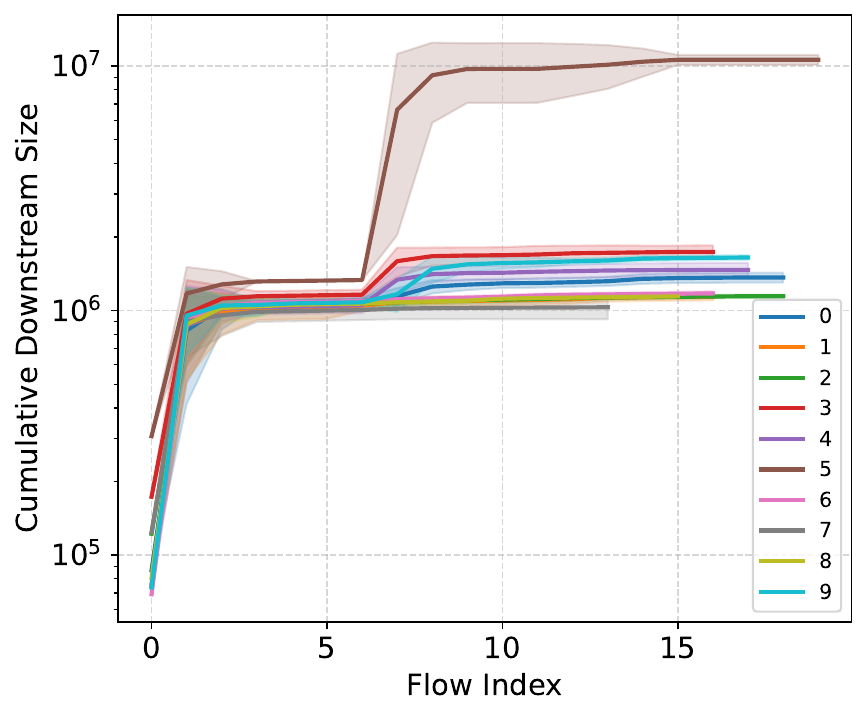}
    \label{subfig:github_feature_down}
}
\hfill
\caption{Visualization Comparison between Packet Length Sequence and UPFS.}
\label{fig:divs}
\end{figure*}

\begin{figure*}[htbp]
\centering
\subfloat[Github Closed]{
\includegraphics[width=0.23\linewidth]{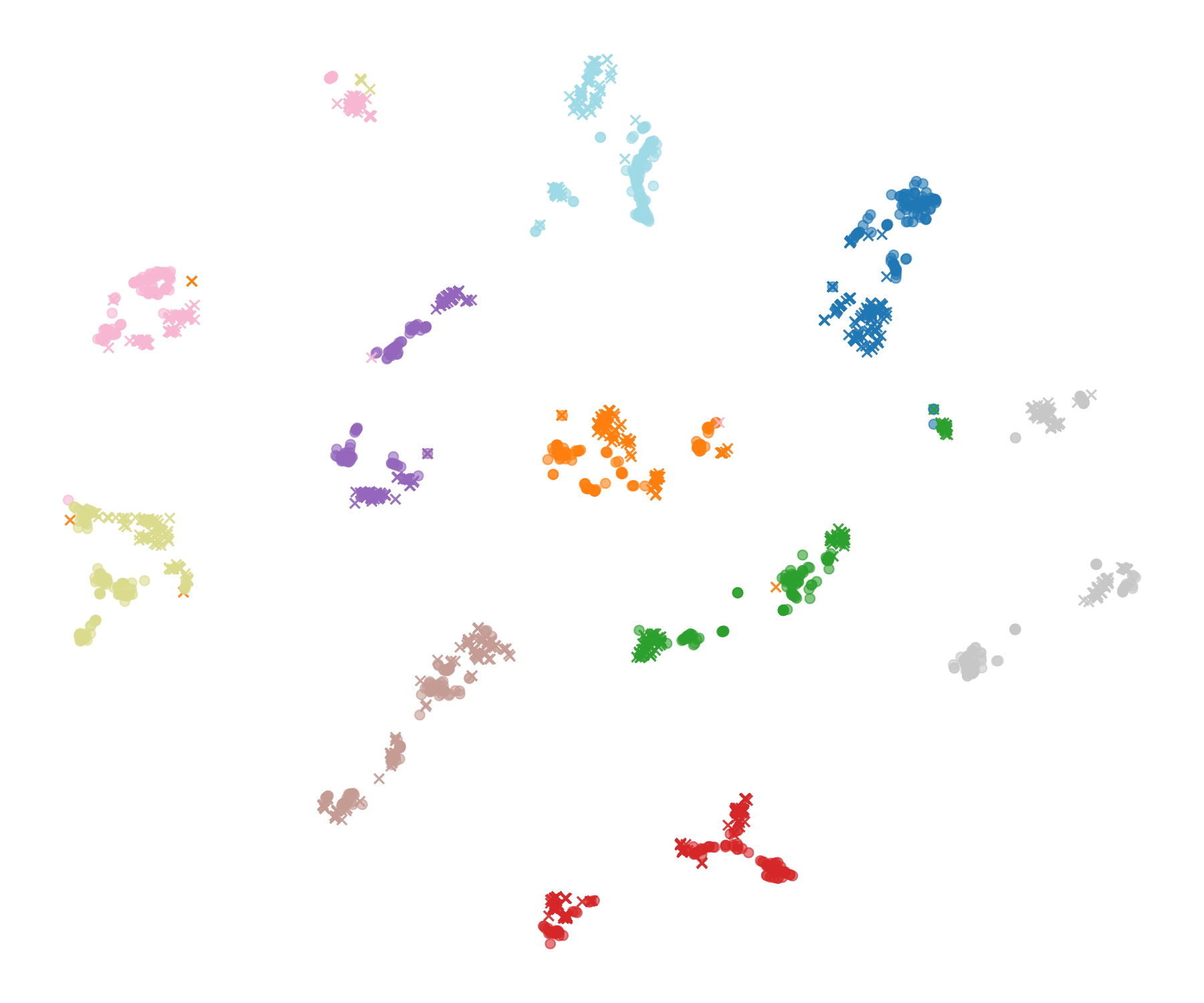}
    \label{subfig:github_tsne_closed_set}
}
\hfill
\subfloat[Github Open-World]{
\includegraphics[width=0.23\linewidth]{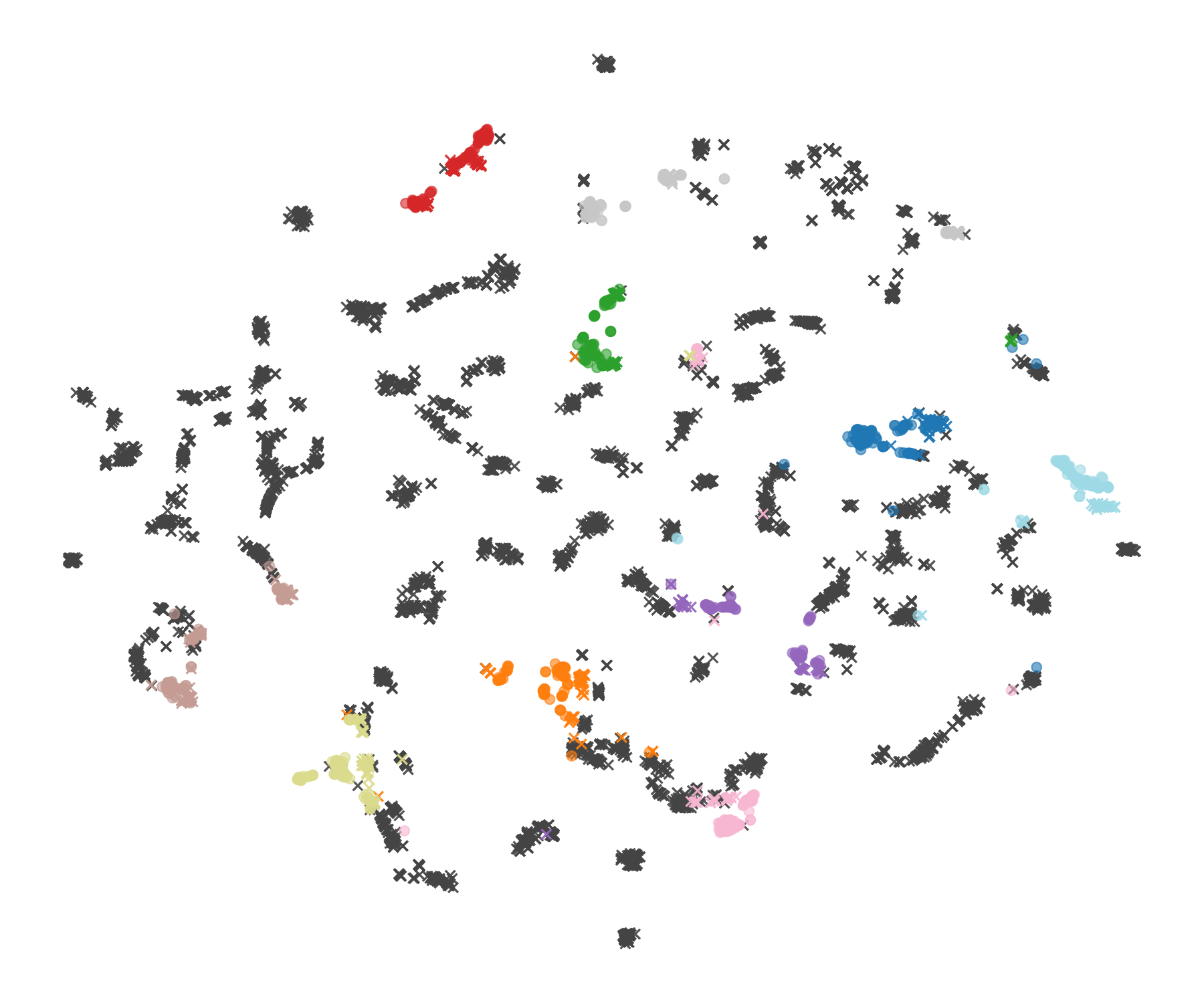}
    \label{subfig:github_tsne_open_set}
}
\hfill
\subfloat[Baidu Closed]{
\includegraphics[width=0.23\linewidth]{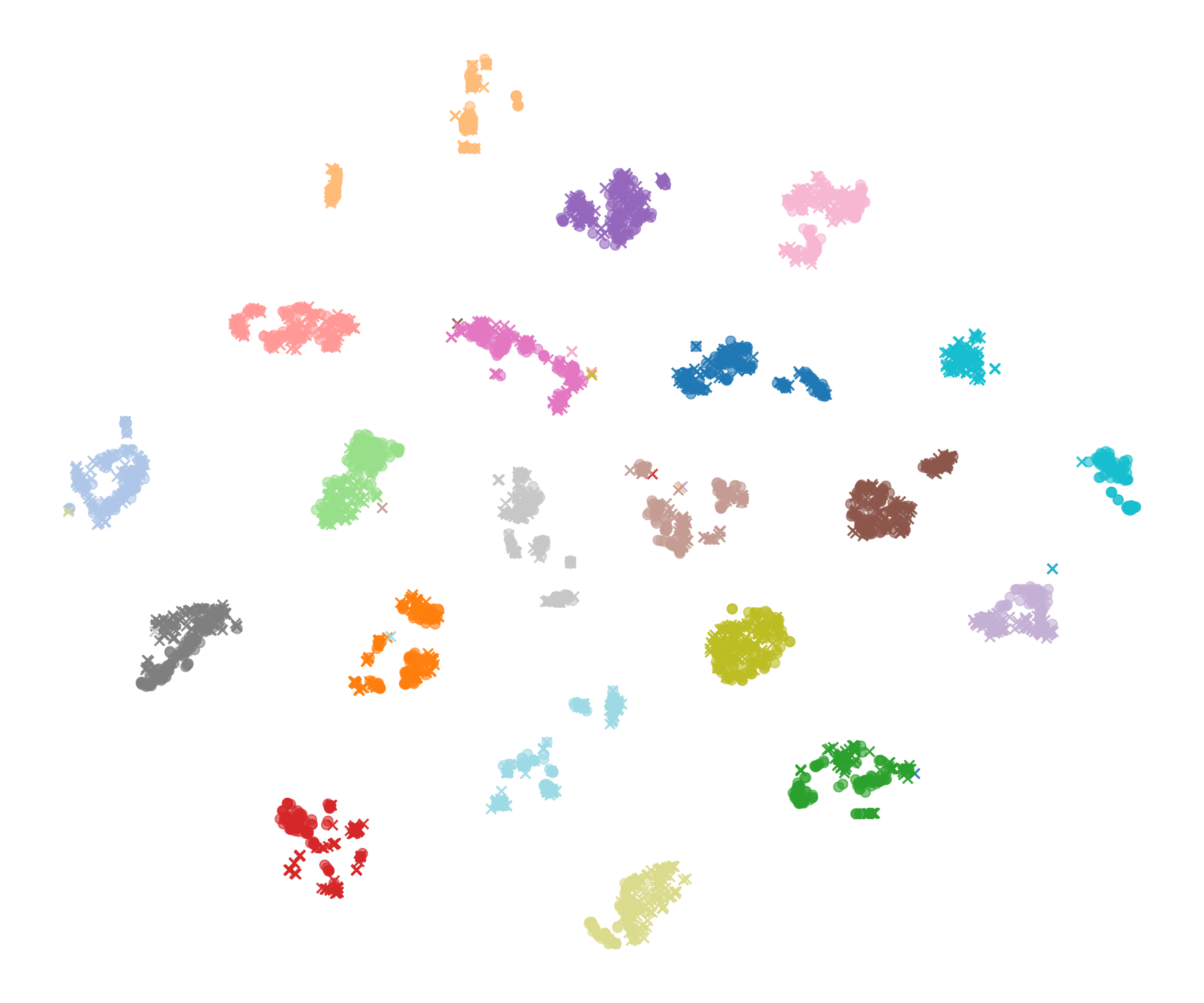}
    \label{subfig:baidu_tsne_closed_set}
}
\hfill
\subfloat[Baidu Open-World]{
\includegraphics[width=0.23\linewidth]{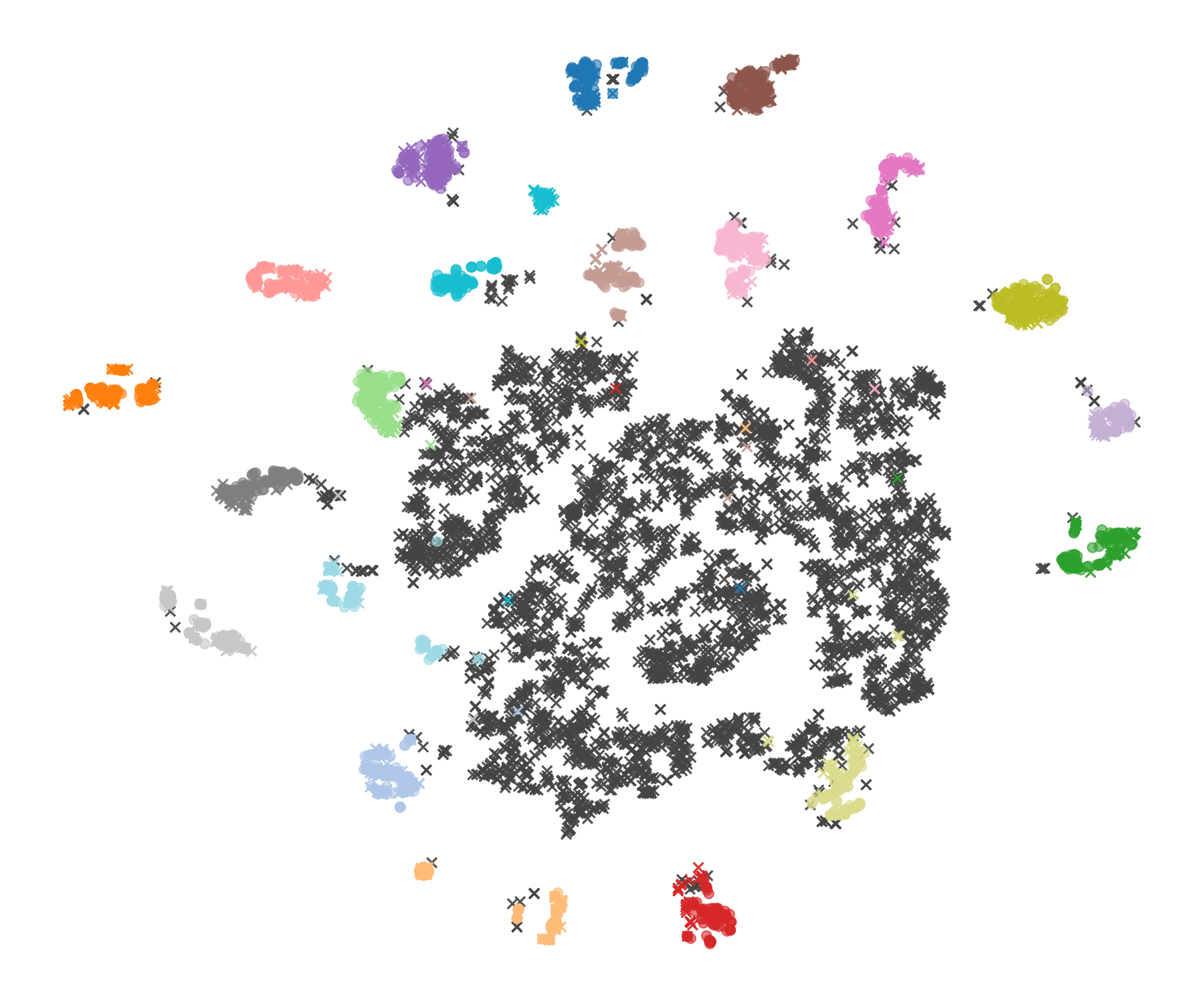}
    \label{subfig:baidu_tsne_open_set}
}
\caption{T-SNE visualization shows the clustering effect of the proposed method under concept drift and open-world scenarios. Different colors represent different classes, circles denote training samples, × marks denote testing samples, and gray × marks denote unknown class samples.}
\label{fig:tsne}
\end{figure*}

\paragraph{Effectiveness Analysis of UPFS}
UPFS is designed to extract efficient and discriminative feature representations from lengthy raw traces. Its effectiveness is reflected in the following three aspects:
\begin{enumerate}
    \item \textbf{Performance Contribution}: As shown in Table \ref{tab:closed_set} and Table \ref{tab:open_set}, removing the UPFS module (`w/o UPFS') and using packet length sequences as feature inputs leads to a significant performance drop. On the most challenging GitHub dataset, the F1 score in the concept drift scenario drops from 97.63\% to 20.33\%, and in the open-world scenario from 54.85\% to 20.23\%. This clearly demonstrates that UPFS is not a simple sequence truncation, but rather successfully extracts key semantic features essential for handling complex scenarios, and is core to maintaining high performance.
    \item \textbf{Feature Quality}: As shown in Figures \ref{fig:divs} and \ref{fig:tsne}, we visually analyzed both the original UPFS and the features learned by the model. The results show that, first, UPFS maintains visually distinguishable patterns compared to the lengthy Trace-level packet length sequences. Second, whether in closed-set or open-world scenarios, the traffic samples of different classes extracted by UPFS form clearly separated and highly distinctive clusters in the feature space. It demonstrates that UPFS can effectively capture the intrinsic differences between application traces, providing high-quality feature inputs for downstream tasks.
    \item \textbf{Temporal and Computational Efficiency}: As shown in Table \ref{tab:time_comparison}, UPFS brings significant efficiency gains. When processing a single Trace sample, our method achieves training and inference times (e.g., 0.2986 ms / 0.0530 ms on the GitHub dataset) comparable to those of the lightweight DF model (0.2622 ms / 0.0374 ms), but far faster than the attention-based TraceClassifier. Notably, without UPFS (`w/o UPFS'), due to the need to process full-length sequences, the per-sample training time rises sharply to 47.5 ms—over 150× slower than Ours. More importantly, UPFS significantly compresses the input sequence length: compared to DF, ours reduces the length by an average of 76×; compared to TraceClassifier, ours reduces it by an average of 15.8×. These results clearly demonstrate that UPFS generates compact and efficient feature representations, greatly reducing computational and storage overhead while improving recognition accuracy, thereby substantially enhancing spatiotemporal efficiency.
\end{enumerate}

\begin{table}[htbp]
\centering
\caption{Comparison of Average Per-Sample Training/Testing Time and Input Sequence Length.}
\label{tab:time_comparison}
\renewcommand{\arraystretch}{1.2}
\resizebox{\linewidth}{!}{ %
\begin{tabular}{lcccc}
\toprule
Method & Train Time(ms) & Test Time(ms) & Sample Unit & Train Set Length \\
\midrule
ET-BERT	&	5.4193	& 1.6631 & Packet & 128 \\
FS-Net           & 0.280   & 0.1593 & Flow & 5000  \\ \midrule
DF               & 0.2622  & 0.0374 & Trace & 5000 \\
TraceClassifier (GitHub) & 20.4722 & 19.9857 & Trace & 731.843 \\
TraceClassifier (Baidu) & 170.7782 & 50.1332 & Trace & 2453.1165 \\
w/o UPFS &	47.502	& 0.6147 & Trace & 5000 \\
Ours (GitHub)   & 0.2986   & 0.0530 & Trace & \textbf{39.98} \\
Ours (Baidu)    & 0.3679   & 0.0565 & Trace & \textbf{184.63} \\
\bottomrule
\end{tabular}
}
\end{table}

\paragraph{Effectiveness of the Open-World Recognition Mechanism}
To validate the effectiveness of the proposed open-world recognition mechanism, we conducted ablation experiments, with results shown in Table \ref{tab:closed_set} and Table \ref{tab:open_set}.
First, the prototypical network is the cornerstone of the framework. Replacing it with a traditional classifier (`w/o Prototype') causes the F1 score in the GitHub closed-set task to drop from 97.99\% to 84.92\%, a significant performance decline. Retaining only the prototypical network (`w/o Weight') still maintains high performance, indicating that the prototypical network is indispensable for constructing a discriminative feature space.
Furthermore, dedicated components play a key role in open-world tasks. Removing the weighted loss (`w/o Weighted Loss') decreases the F1 score from 54.85\% to 45.28\%, showing that without this constraint, feature distributions become loose and distinguishing unknown classes becomes difficult. Removing the adaptive weighted threshold (`w/o Weighted Threshold') further decreases performance to 44.00\%, verifying the necessity of class-specific decision boundaries in fine-grained tasks.
In summary, the prototypical network provides a solid foundation for feature representation, while the weighted loss and adaptive threshold improve the model’s ability to recognize unknown classes during training and inference, respectively. The three components together ensure the overall effectiveness of the proposed method.

\begin{mybox}
\textbf{Answer for RQ2: } The ablation study results strongly demonstrate that UPFS is an efficient and highly discriminative feature representation method, and our designed open-world recognition mechanism robustly handles unknown traffic. Together, they constitute the key to the success of our framework.
\end{mybox}

\subsection{Robustness Analysis (Q3)}
 \label{sec:sensitivity_analysis}
\begin{figure}[h]
    \centering
    \includegraphics[width=0.8\linewidth]{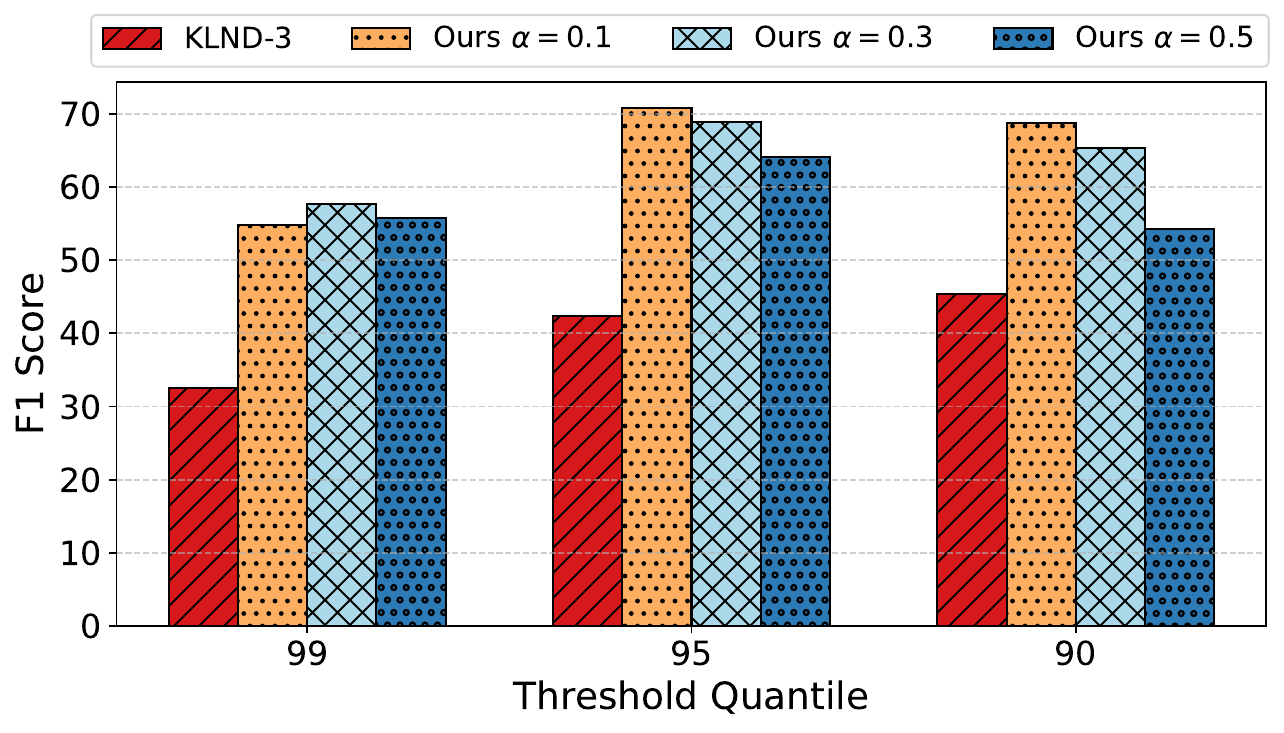}
    \caption{Parameter sensitivity analysis.}
    \label{fig:robust_thresholds}
\end{figure}

\begin{figure}[h]
    \centering
    \includegraphics[width=0.8\linewidth]{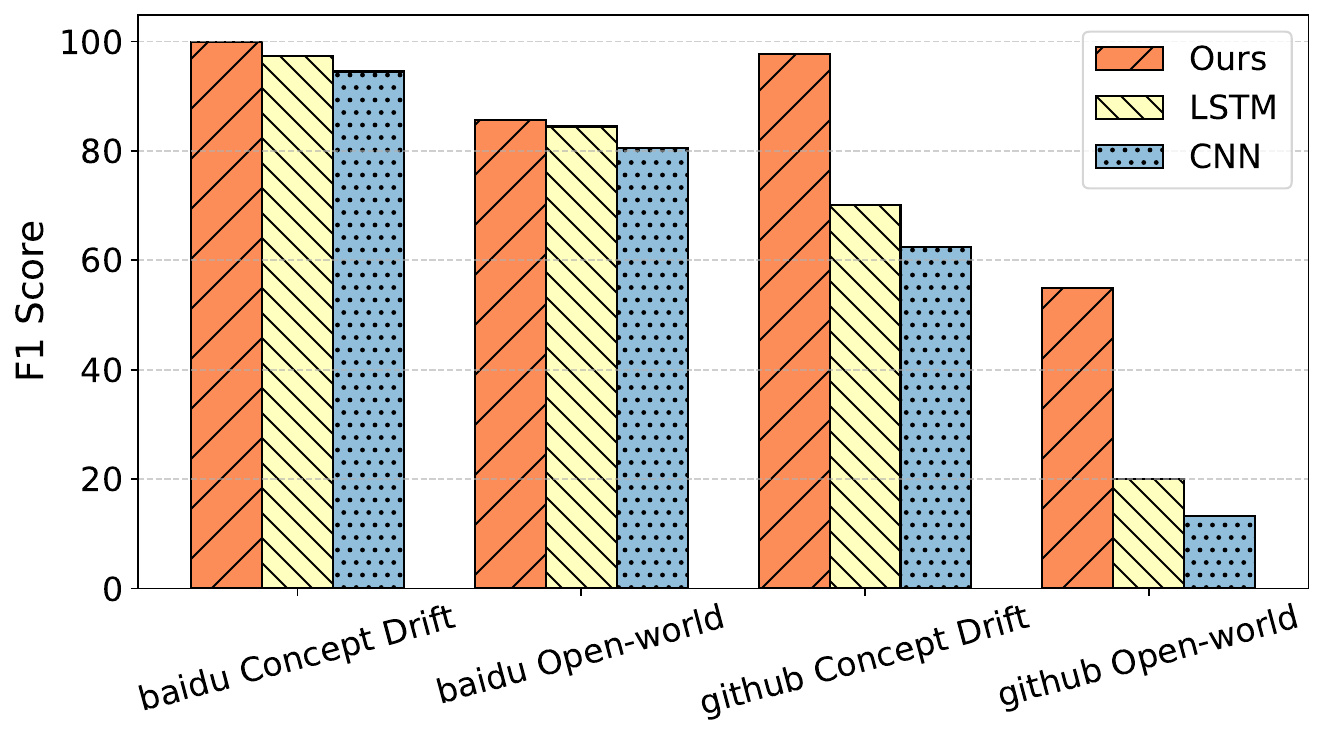}
    \caption{Performance comparison of different feature extractors.}
    \label{fig:robust_models}
\end{figure}

This section aims to investigate the robustness of the proposed method from two perspectives: the sensitivity of key hyperparameters and the generalizability of the overall framework, to address Research Question Q3. All experiments are conducted on the most challenging fine-grained GitHub dataset.
\paragraph{Hyperparameter Sensitivity Analysis}
 We first evaluate the impact of different threshold percentiles~$p$ and weighting coefficient~$\alpha$ on open-world recognition performance. As shown in Fig.~\ref{fig:robust_thresholds}, the results clearly indicate that across all tested percentiles $p \in [90, 95, 99]$, our method consistently achieves significantly higher F1 scores than the baseline model KLND-3, regardless of the value of $\alpha$. This demonstrates that our approach exhibits strong robustness to hyperparameter selection—achieving stable and reliable performance without the need for meticulous tuning—showcasing its strong potential for practical deployment.
\paragraph{Generalizability Analysis}
 To further verify the universality and modular flexibility of the proposed framework, we replace the original Transformer-based feature extractor with typical architectures such as LSTM and CNN, while keeping the overall pipeline unchanged. As shown in Fig.~\ref{fig:robust_models}, for the fine-grained GitHub task that requires capturing subtle behavioral variations, replacing the Transformer with a weaker feature extractor (e.g., CNN) leads to a noticeable performance drop. In contrast, for the coarse-grained Baidu task where inter-class differences are more pronounced, the performance degradation is relatively minor. These results not only validate the general applicability of the proposed method across different tasks but also highlight its modular advantage—users can flexibly select feature extractors of varying complexity to achieve an optimal balance between performance and computational efficiency.

\begin{mybox}
 \textbf{Answer for RQ3: }
 The proposed method demonstrates low parameter sensitivity and strong architectural robustness. Key hyperparameters do not require fine-tuning to achieve superior performance, and the core framework is generalizable, allowing integration with various backbone networks as needed. These characteristics ensure high stability and portability of the method in practical deployments.
 \end{mybox}

\section{Conclusion}
\label{conclusion}
To address the accuracy–overhead trade-off faced by existing sequence-based features, this paper proposes a novel encrypted traffic representation method — UPFS. UPFS efficiently compresses the length of traffic sequences while preserving critical discriminative features, significantly improving representation efficiency. Based on this, we design an adaptive open-world recognition mechanism built upon prototype networks, which learns personalized decision boundaries for each class to enhance classification performance and effectively reject unknown traffic. Extensive experiments across various complex scenarios, including open-world settings, demonstrate that the proposed method significantly outperforms existing approaches in both accuracy and generalization, validating its effectiveness and applicability in real-world network environments.

In the future, our research will focus on two key challenges: first, overcoming feature-space collision in fine-grained scenarios by employing more advanced representation techniques to enhance the model’s ability to recognize concept drift and similar unknown classes; second, improving model interpretability by revealing critical traffic features, thereby transforming the black box into a transparent and trustworthy diagnostic tool, and enhancing its applicability in security-critical scenarios.

\bibliographystyle{IEEEtran}
\bibliography{IEEEabrv,ref}

\begin{IEEEbiography}[{\includegraphics[width=1in,height=1.25in,clip,keepaspectratio]{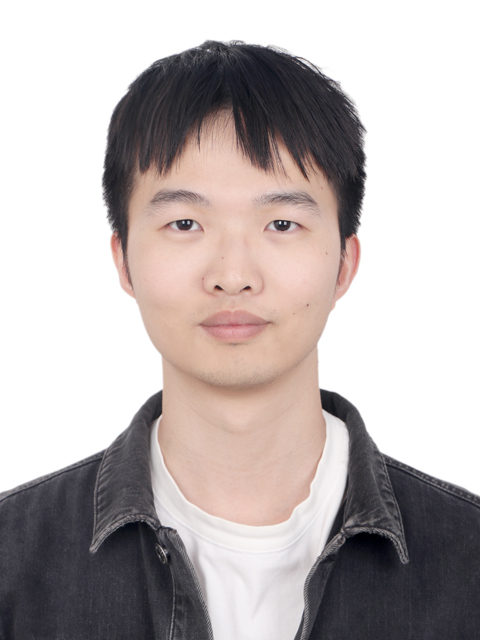}}]{Youquan Xian}
 obtained his master's degree from Guangxi Normal University in 2024 and is currently pursuing a doctoral degree at Beijing University of Posts and Telecommunications. He has published over 10 papers in journals and conferences, including IEEE Transactions on Parallel and Distributed Systems, ACM Transactions on the Web, IEEE Transactions on Network and Service Management, Future Generation Computer Systems, and ASPLOS 2025. His primary research focuses on Network Security, Blockchain, and Federated Learning.\end{IEEEbiography}

\begin{IEEEbiography}[{\includegraphics[width=1in,height=1.25in,clip,keepaspectratio]{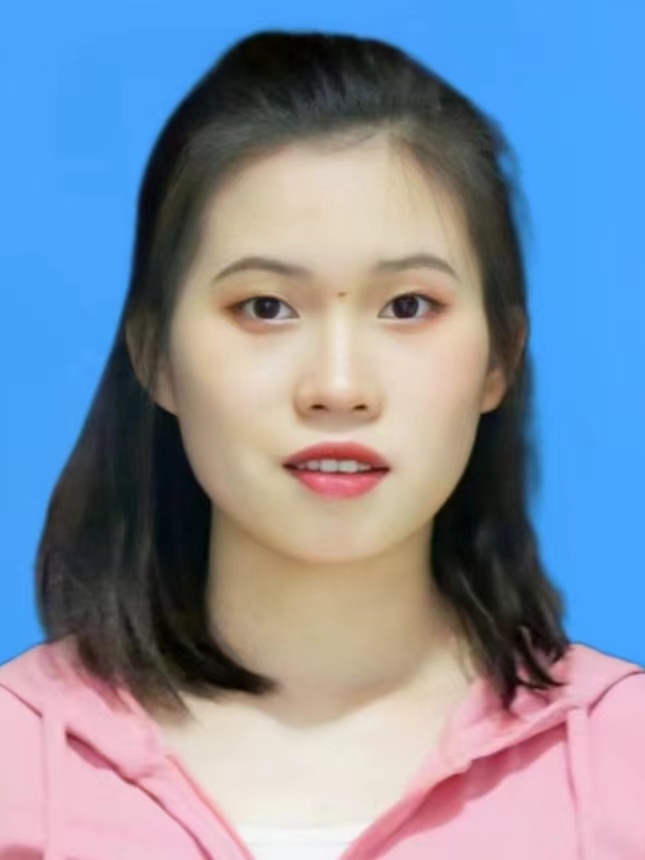}}]{Xueying Zeng}
  obtained his master's degree from Guangxi Normal University in 2025 and is currently pursuing a doctoral degree at Beihang University. She has published over 5 papers in journals and conferences, including ACM Transactions on the Web, IEEE Transactions on Parallel and Distributed Systems, Future Generation Computer Systems, and IEEE SMC 2024. Her main research areas are Software Security and Blockchain.\end{IEEEbiography}

\begin{IEEEbiography}[{\includegraphics[width=1in,height=1.25in,clip,keepaspectratio]{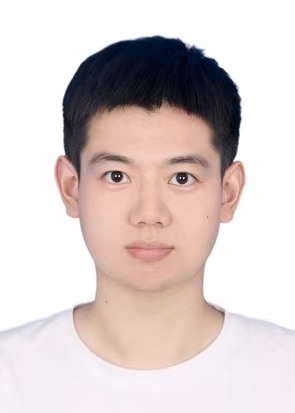}}]{Aoxiang Zhou}
is a master's student at the School of Computer Science and Engineering, Guangxi Normal University. His main research interests include natural language processing and federated learning.
\end{IEEEbiography}

\begin{IEEEbiography}[{\includegraphics[width=1in,height=1.25in,clip,keepaspectratio]{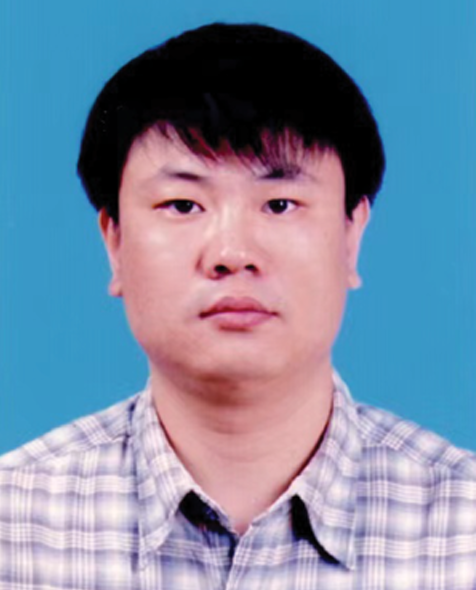}}]{Jinqiao Shi}
 was born in 1978. He received the Ph.D. degree from Harbin Institute of Technology, Harbin, China. He is currently a Professor and a Ph.D. Supervisor with Beijing University of Posts and Telecommunications, Beijing, China. His current research interests include network privacy and anonymity, artificial intelligence security, and traffic analysis.
\end{IEEEbiography}

\begin{IEEEbiography}[{\includegraphics[width=1in,height=1.25in,clip,keepaspectratio]{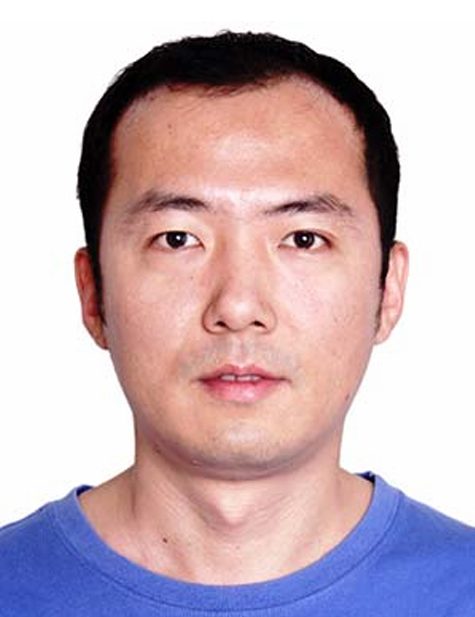}}]{Zhiyu Hao}
 received the doctor’s degree in computer system architecture from the Harbin Institute of Technology, in 2007. He is currently a professor with Zhongguancun Laboratory, Beijing. His research interests include network security, system virtualization. He has published more than 40 papers in journals
 and conferences including the IEEE Transactions on
 Parallel and Distributed Systems, ICPP, IEEE Secu
rity \& Privacy, ICA3PP and CLUSTER.
\end{IEEEbiography}

\begin{IEEEbiography}[{\includegraphics[width=1in,height=1.25in,clip,keepaspectratio]{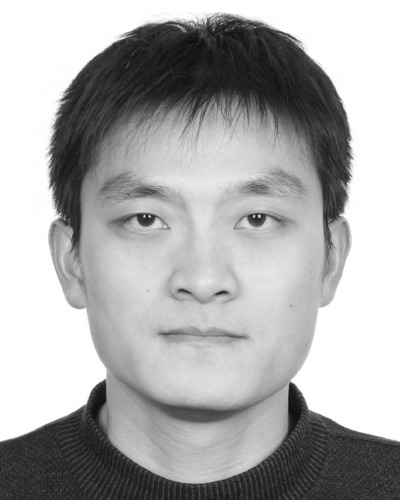}}]{Lei Cui}
 received the Ph.D. degree in computer
 software and theory from Beihang University
 in 2015. He is currently an Associate Professor
 with the Zhongguancun Laboratory, Beijing. He
 has published over 40 papers in journals and
 conferences, including IEEE TRANSACTIONS ON
 PARALLEL AND DISTRIBUTED SYSTEMS, IEEE
 TRANSACTIONS ON INFORMATION FORENSICS
 AND SECURITY, IEEE TRANSACTIONS ON
 SERVICES COMPUTING, ISSTA, ICCD, RAID,
 VEE, LISA, and DSN. His research interests
 include operating systems, system security, and system virtualization.
\end{IEEEbiography}

\begin{IEEEbiography}[{\includegraphics[width=1in,height=1.25in,clip,keepaspectratio]{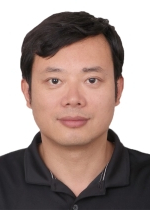}}]{Peng Liu}
received his Ph.D. degree in 2017 from Beihang University, China. He began his academic career as an assistant professor at Guangxi Normal University in 2007 and was promoted to full professor in 2022. His current research interests are focused on federated learning and blockchain.
\end{IEEEbiography}

\end{document}